\documentclass[10pt,preprint]{aastex}

\shorttitle{Multi-timescale of QPPs} \shortauthors{Tan et al.}

\begin{document}

\title{Microwave Quasi-periodic Pulsations in Multi-timescales Associated with a Solar Flare/CME Event}
\author{Baolin Tan\altaffilmark{1}, Yin Zhang, Chengming Tan, \& Yuying Liu}
\affil{Key Laboratory of Solar Activity, National Astronomical
Observatories \\ Chinese Academy of Sciences, China}
\email{bltan@nao.cas.cn}

\altaffiltext{1}{Datun Road A20, Chaoyang District, Beijing,
100012, China.}

\begin{abstract}

Microwave observations of quasi-periodic pulsations (QPP) in
multi-timescales are confirmed to be associated with an X3.4
flare/CME event at Solar Broadband Radio Spectrometer in Huairou
(SBRS/Huairou) on 13 December 2006. It is most remarkable that the
timescales of QPPs are distributed in a broad range from
hecto-second (very long period pulsation, VLP, the period $P>100$
s), deca-second (long period pulsation, LPP, $10<P<100$ s), few
seconds (short period pulsation, SPP, $1<P<10$ s), deci-second
(slow-very short period pulsation, slow-VSP, $0.1<P<1.0$ s), to
centi-second (fast-very short period pulsation, fast-VSP, $P<0.1$
s), and forms a broad hierarchy of timescales. The statistical
distribution in logarithmic period-duration space indicates that
QPPs can be classified into two groups: group I includes VLP, LPP,
SPP and part of slow-VSPs distributed around a line approximately;
group II includes fast-VSP and most of slow-VSP dispersively
distributed away from the above line. This feature implies that
the generation mechanism of group I is different from group II.
Group I is possibly related with some MHD oscillations in
magnetized plasma loops in the active region, e.g., VLP may be
generated by standing slow sausage mode coupling and resonating
with the underlying photospheric 5-min oscillation, the modulation
is amplified and forms the main framework of the whole flare/CME
process; LPP, SPP, and part of slow-VSPs are most likely to be
caused by standing fast modes or LRC-circuit resonance in
current-carrying plasma loops. Group II is possibly generated by
modulations of resistive tearing-mode oscillations in electric
current-carrying flaring loops.

\end{abstract}


\section{Introduction}

Solar quasi-periodic pulsations (QPP) are frequently observed in
optical, EUV, radio, soft X-ray, hard X-ray, and even Gamma-ray
emissions (Nakariakov et al, 2010). Same as seismological waves
can reveal interior structure of the Earth or other celestial
bodies, solar QPPs are also a kind of very important phenomena
which may provide some information, such as the interior
structures and the physical conditions of the source regions or
the propagating media. The flare-associated QPP can provide
information of solar flaring regions, and give some prospective
enlightening of coronal plasma dynamic processes, such as to
remote diagnose the microphysics of energy releasing sites.
Especially the QPP occurred in microwave frequency range is always
regarded as a direct signal of flaring primary energy-releasing
regions. So far, from observations we know the timescales
(expressed as period $P$) of QPPs are in a wide range from a few
tens of milliseconds to several minutes. Based on the summary of
previous observational and theoretical investigations, according
to the timescale of periods, we may extend the previous QPP
classification (Wang \& Xie, 2000) into a much wider hierarchy:

(1) Very long period pulsation (VLP), which is also named as very
low frequency oscillation. Its period is in hecto-second or
several minutes. Generally we may define VLP as $P>100$ s (e.g. 40
min, Kaufmann, 1972; 276 s, Aschwanden et al, 1999; 8 -- 12 min,
Foullon et al, 2005; 10-18 min, Foullon et al, 2010).

(2) Long period pulsation (LPP). The period is in deca-second,
defined as $10 < P < 100$ s (e.g. Parks \& Winckler, 1969; Qin et
al, 1996; Melnikov et al, 2005; Inglis \& Nakariakov, 2009; etc).

(3) Short period pulsation (SPP). This kind of QPP is very
popular, and its period is in seconds, $1 \leq P < 10 $ s (e.g.
Abrami, 1970; Rosenberg, 1970; Trottet et al, 1981; Zaitsev \&
Stepanov, 1982; Qin et al, 1996; etc).

(4) Very short period pulsation (VSP). Its period is in
sub-second. Frequently it is always down to tens of milliseconds
(e.g. Young et al, 1961; Gotwols, 1972; Fleishman et al, 2002; Tan
et al, 2007; Tan, 2008; Jiricka \& Karlicky, 2008; Karlicky,
Zlobec, \& Meszarosova, 2010; etc), so we may classify VSPs into
two sub-classes (Tan et al, 2007):

1) slow-VSP, the period is in deci-second, $0.1<P<1.0$ s;

2) fast-VSP, the period is in centi-second, $P<0.1$ s, usually, it
is only several tens of millisecond.

Intrinsically, there is no distinct borderline between different
classes of QPPs. So far, because of instrumental limitations, we
have not distinguished any QPPs with period of shorter than 10
milliseconds reliably.

In previous studies, there is always only one or two classes of
QPPs reported in one flare event. Even if the multi-periodic
pulsations observed in some cases, they are also only with one or
two classes of QPPs. For example, in the work of Qin et al (1996),
the periods of the two pulsating components are 1.5 s and 40 s
which are belonging to SPP and LPP, respectively; in the work of
Melnikov et al (2005), the periods of two pulsating components are
14 -- 17 s and 8 -- 11 s, respectively, which are ranked in same
class of QPP; in the work of Inglis \& Nakariakov (2009), the
periods of three pulsating components are 28 s, 18 s, and 12 s,
respectively, and all of them are belonging to one and the same
class of LPP. Karlicky et al (2005) reported the timescales of
slowly drifting pulsating structures (DPS) with period of 0.9--7.5
s, and the short periods presents as a power-law distribution,
especially in the range of 0.06-0.2 s, where the power-low index
is in the range of 1.3-1.6. So far we have no literatures to show
the coexistence of more than two different classes of QPP in a
same flaring event. We do not know what is the relationship among
different classes of QPPs. At the same time we have plenty of
reasons to suppose that such relationships may imply some physical
information of the solar active region or the flaring mechanisms.

It is very luck that an X3.4 flare/CME event occurred in solar
active region of AR 10930 on 2006 December 13. This event has many
unusual features: (1) AR 10930 is an isolated active region on the
solar disk (Left panel of Fig.1); (2) the flare is a long duration
event, the GOES soft X-ray bursts start at 02:14 UT and end at
02:57 UT, and the radio bursts start at 02:20 UT and last to after
04:50 UT; (3) the eruption is repetitious, from broadband
microwave observations, we may find that this event has many big
bursts clearly. Many people studied this flare/CME event from
different points of view (Kosovichev \& Sekii, 2007; Yan, et al,
2007; Ning, 2008; Kuznetsev, 2008; Zhang et al, 2008; etc).
Especially, in the work of Minoshima et al (2009), they presented
multi-wavelength observations of electron acceleration in the
flare. And in the Fig.1 of their paper, the light curves of
microwave radio emission observed at NoRP (9.4 GHz, 17 GHz, and 34
GHz) and hard X-ray obtained at RHESSI in 25 -- 40 keV, 40 -- 60
keV demonstrate quasi-periodic pulsations with very long period.
Together with the great number of very short period pulsations
(VSP) observed in Chinese Solar Broadband Radio Spectrometer
(SBRS/Huairou) (Tan et al, 2007; Tan, 2008), it is rational to
investigate the observations of QPPs in multi-timescales, their
mutual relationships, and the possible physical implications in
detail. From these investigations, we find that the most
remarkable feature in this flare event is the coexistence of
several classes of QPPs, including VLP, LPP, SPP, slow-VSP and
fast-VSP, and these QPPs form a broad hierarchy of timescales.

In this work, based on the observations at frequency of 2.60 --
3.80 GHz at SBRS/Huairou, we present the observations and analysis
of multi-timescales in QPPs associated with the flare event on
2006 December 13. Section 2 is an introduction of the observation
data and analysis methods. Section 3 presents the main features of
QPPs. In section 4 we give a detailed discussion of physical
mechanisms of QPPs with multi-timescales. Finally, there is a
summary and some deductions in section 5.

\section{Observations Data and Analysis}

\subsection{Observation Data}

The X3.4 flare/CME event in AR 10930 on 2006 December 13 is
observed by RHESSI, Hinode, NoRH, NoRP, SOHO/MDI, TRACE and
SBRS/Huairou, etc (Kubo et al, 2007; Su et al, 2007; et al). We
select mainly the observations of SBRS/Huairou because of its high
cadence, broad frequency bandwidth, and high frequency resolution,
to investigate the multi-timescale pulsating phenomena associated
with the flare/CME event. SBRS/Huairou includes 3 parts: 1.10-2.06
GHz, 2.60-3.80 GHz, and 5.20-7.60 GHz (Fu et al 1995; Fu et al
2004; Yan et al, 2002). However, only the spectrometer of 2.60 --
3.80 GHz is operating well around the above flare/CME event. The
diameter of the antenna of the spectrometer is 3.2 m. The antenna
points to the center of solar disk automatically controlled by a
computer. The spectrometer can receive the total flux of solar
radio emission with dual circular polarization (left- and right
circular polarization), and the dynamic range is 10 dB above quiet
solar background emission. And the observation sensitivity is:
$S/S_{\bigodot}\leq 2\%$, here $S_{\bigodot}$ is quiet solar
background emission. Similar to other several spectrometers, such
as Phoenix (100 -- 4000 MHz, Benz et al, 1991), Ond\'rejov (800 --
4500 MHz, Jiricka et al, 1993) and BBS (200 -- 2500 MHz, Sawant et
al, 2001), SBRS/Huairou have no spatial resolution. However, a
great deal of works (e.g. Dulk, 1985, etc) show that the radio
bursts received by spectrometers are always coming from the solar
active region when the antenna points to the Sun. In this work,
because AR 10930 is an isolated active region around the X3.4
flare/CME event (shows in the left panel of Fig.1), we believe
that the microwave bursts are just coming from the same active
region associated with the event. In the frequency range of 2.60 -
3.80 GHz, there are 120 channels with frequency resolution of 10
MHz, temporal resolution of 8 ms, left- and right- circular
polarization components with accuracy of polarization degree 5 --
10 \%. The data were analyzed by using a software that was
developed with IDL algorithm. In order to identify weak burst
structures, some wavelet methods have been developed for data
processing. The calibration of the observation data is followed a
method proposed by Tanaka et al (1973). The standard flux values
of the quiet Sun is adopted the data published by Solar
Geophysical Data ($SGD$) at frequencies 4995 MHz, 2800 MHz, 2695
Mhz and 1415 Mhz. As for strong bursts, the receiver may work
beyond its linear range and a nonlinear calibration method will be
used instead (Yan et al, 2002).

\begin{figure}
\begin{center}
\includegraphics[width=6.5cm]{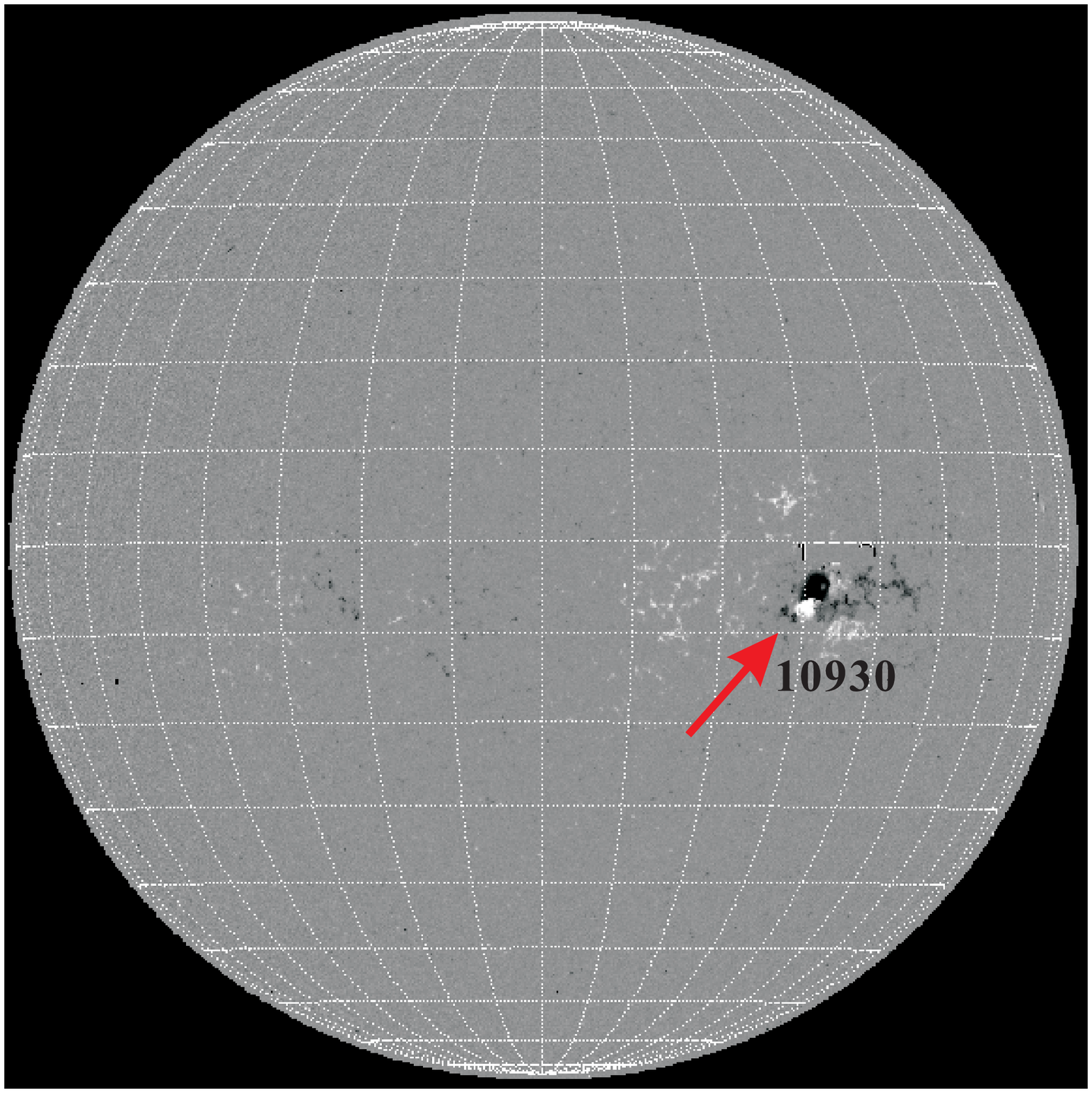}
\includegraphics[width=9.2cm]{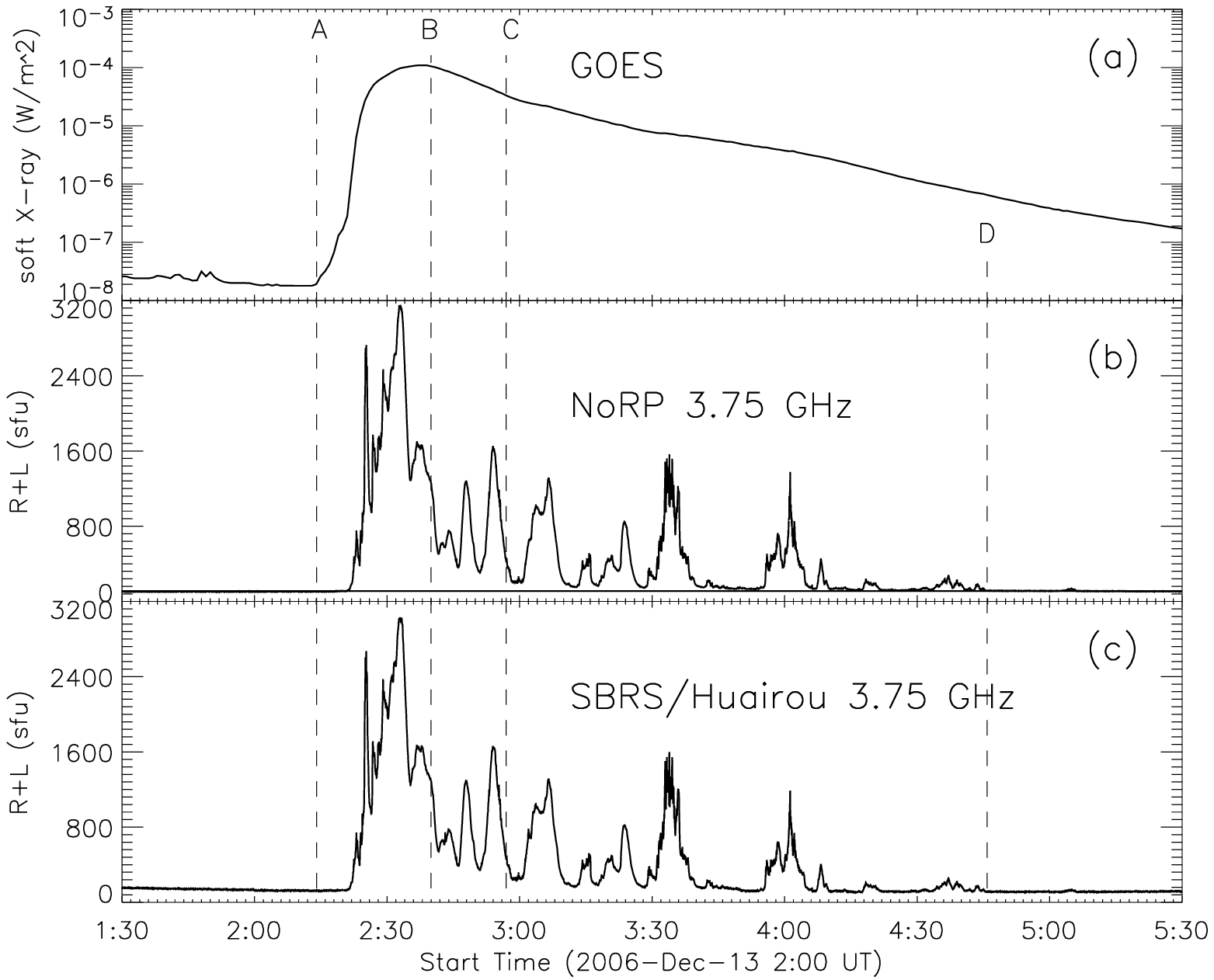}
\caption{Left panel is a full-disk longitudinal magnetogram taken
by MDI/SOHO on 13 Dec. 2006. The right panel presents observation
comparisons between solar soft X-ray intensity taken by GOES
satellite (a), radio emission intensities at frequency of 3.75 GHz
taken by NoRP (b, in Japan) and SBRS/Huairou (c, in China) at
01:30 -- 05:30 UT, 13 Dec. 2006. The dashed vertical lines marked
as A, B, C and D indicate the start, peak, end time of the soft
X-ray flare observed by the GOES, and the end time of the radio
burst, respectively.}
\end{center}
\end{figure}

\begin{figure}
\begin{center}
\includegraphics[width=8.2cm]{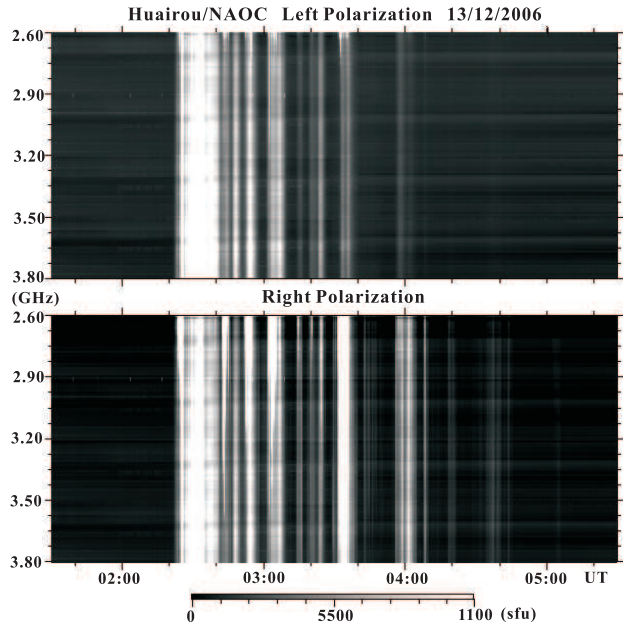}
\caption{Spectrogram of the solar radio emission at frequency of
2.60 -- 3.80 GHz at 01:30 -- 05:30 UT, 13 Dec. 2006.}
\end{center}
\end{figure}

In order to confirm that the observations of SBRS/Huairou are not
artificial, we make comparisons among SBRS/Huairou, Nobeyama Radio
Polarimeters (NoRP) and soft X-ray intensity obtained at GOES
satellite. Here, NoRP is observing the Sun with multiple
frequencies 1.0. 2.0, 3.75, 9.4, 80 GHz. It is capable to obtain
solar total flux and circular-polarization degree with temporal
resolution of 0.2 s. The right panel of Fig.1 presents the
observed results of GOES soft X-ray intensity (a), radio emission
intensities obtained at NoRP (b) and SBRS/Huairou (c) at the same
frequency of 3.75 GHz at 01:30 -- 05:30 UT, 13 Dec. 2006. We find
that these radio results obtained from the two different
telescopes (one is in China, and the other is in Japan) are almost
selfsame in profile shape and intensity. Fig.2 presents the
dynamic spectrogram of solar radio emission at frequency of 2.60
-- 3.80 GHz at 01:30 -- 05:30 UT, 13 Dec. 2006. Radio bursts start
(02:20 UT) after the flare onset (02:14 UT), and end at 04:50 UT.
They are consistent with the enhancement of GOES soft X-ray
intensity. These facts confirm again that radio observations of
SBRS/Huairou are coming from the solar flaring processes.

The most advantage of SBRS/Huairou is that it is possible to
provide much of peculiar information synchronously, such as
polarization degree, pulsating frequency bandwidth, frequency
drifting rate, and fine structures of microwave emission
spectrograms with much higher temporal and frequency resolutions
than most other instruments.

\subsection{Analysis Method}

As a rule, we can use Fourier analysis (Fast Fourier
Transformation, FFT) or wavelet analysis to investigate pulsating
structures in observation data. Generally, QPP always behaves as a
train of approximated paralleled vertical and equidistant stripes
in broadband dynamic spectrograms. Each stripe represents one
pulse. The time interval between two adjacent strips represents
the pulsating period. By scrutinizing observed dynamic
spectrograms, we can also distinguish and confirm the existence of
QPPs clearly, and pickup their parameters easily, such as central
emission frequencies ($f_{0}$), frequency bandwidth ($b_{w}$),
period ($P$), duration ($D$), global frequency drifting rate
($R_{gfd}$), signal pulse frequency drifting rate ($R_{spfd}$),
and polarization degree ($r$), etc. In a QPP, parameters of $P$,
$f_{0}$, $b_{w}$, $r$, $R_{gfd}$ and $R_{spfd}$ may have alike
values. In different QPP, the above parameters will have obviously
different values. We can easily distinguish one QPP from the
others. Then we need to decompose pulsating components from the
observed data. In our observations, the data have two modes
synchronously: one is with cadence of 0.2 s, and the other has
cadence of 8 ms. From observations with cadence of 0.2 s, we may
investigate features of VLP, LPP and part of SPP; and from the
observations with cadence of 8 ms, we may investigate VSP and most
of the SPP.

In order to make the pulsating component more clearly and
reliably, we adopt two methods to investigate QPPS. These two
methods can be cross-examined mutually. When their results are
consistent with each other, a QPP is identified. The first method
is Fourier analysis (Fast Fourier Transformation, FFT) with
observed data. The second method is a kind of statistical method,
which directly counts the temporal intervals between adjacent
pulse peaks in the pulsating structures from the broadband radio
dynamic spectrogram. This method is very straightforward, the
details is presented in following paragraphs.

(1) Identify solar microwave bursts. Because the instrument
sensitivity is $S/S_{\bigodot}\leq 2\%$. From SGD data, we may get
the microwave flux intensity of the quiet solar background
emission at frequency 2.695 GHz, 2.800 GHz, and 4.995 GHz are 52
sfu, 88 sfu, 90.7 sfu, and 158 sfu, respectively, on 2006 December
13. Then we may obtain the microwave flux intensity of quiet solar
background emission at each frequency channel in the range of 2.60
-- 3.80 GHz is 85.3 -- 121.4 sfu by linear extrapolation, and the
instrument sensitivity is about 1.70 -- 2.40 sfu. More securely,
when the microwave flux intensity is 5.0 sfu (more than 2 times
instrument sensitivity in frequency 2.60 -- 3.80 GHz) higher than
the quiet solar background emission, we may confirm the existence
of a solar microwave burst.

(2) Identify pulsating structure by scrutinizing the observed
spectrograms directly. Here, the microwave flux intensity at each
pulse in pulsating structure will be 5.0 sfu higher than the quiet
solar background emission, so it is very clear in dynamic
spectrogram.

(3) Obtain the timescale of pulsating structure approximately by
direct counting the temporal intervals between adjacent pulse
peaks from the spectrogram. Such estimation can give roughly
values of period $P'$, duration $D$, central emission frequency
$f_{0}$, and the bandwidth $b_{w}$ of QPP.

(4) Smooth the observing flux curves to filter out the high
frequency noise components by using a sliding window narrower than
pulsating period, the result may be expressed as $F$. The width of
the smoothing window should be experientially as $\frac{1}{4}P'$.

(5) Smooth the flux curves $F$ in a wide sliding window to filter
out the pulsating component and obtain the background emission of
QPP. The width of the smoothing window should be experientially as
$2P'$. The result can be expressed as $F_{b}$. Generally, the
background emission during the flare/CME event is much greater
than the quiet sun emission (before the event, i.e. before 02:20
UT), and it equals to quiet sun emission before the flare event.

Usually, the smoothing method will generate some additionally
boundary effects. In order to suppress such effects, we need to
extend the range of the analyzing data. That is to say, if the
duration of QPP is $D$, the QPP starts from $t_{1}$ and ends at
$t_{2}$, $D=t_{2}-t_{1}$, then we select the analyzing data from
$t_{1}-\frac{1}{2}D$ to $t_{2}+\frac{1}{2}D$ before make smoothing
processing. After smoothing processing, we investigate the
pulsating features only within the data fragment from $t_{1}$ to
$t_{2}$.

(6) Subtract the background emission $F_{b}$ from the result $F$
of step (4) and obtain the pulsating component ($F_{p}$),
$F_{p}=F-F_{b}$. So, the pulsating component is a quantity
relative to the background emission $F_{b}$, it will oscillate
from positive to negative values around $F_{b}$.

(7) Investigate the dynamic features of QPPs by analyzing $F_{p}$.
Such features may include more exact period $P$, modulation degree
$M$, polarization degree $r$, single pulse frequency drifting
rates $R_{spfd}$, and global frequency drifting rate $R_{gfd}$.
These parameters are defined and explained in paper of Tan (2008).

\begin{figure}
\begin{center}
  \includegraphics[width=8.2cm]{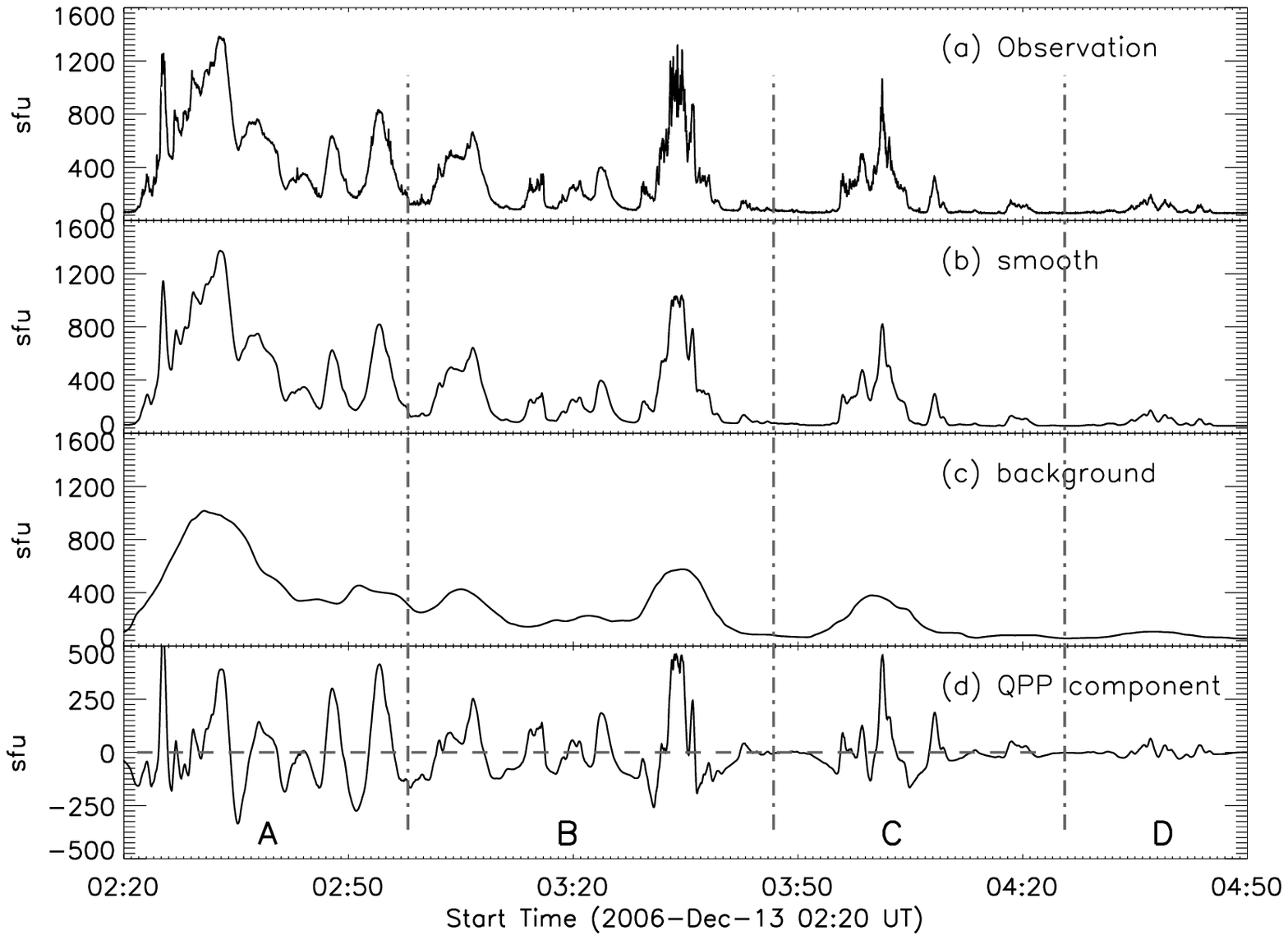}
  \includegraphics[width=8.2cm]{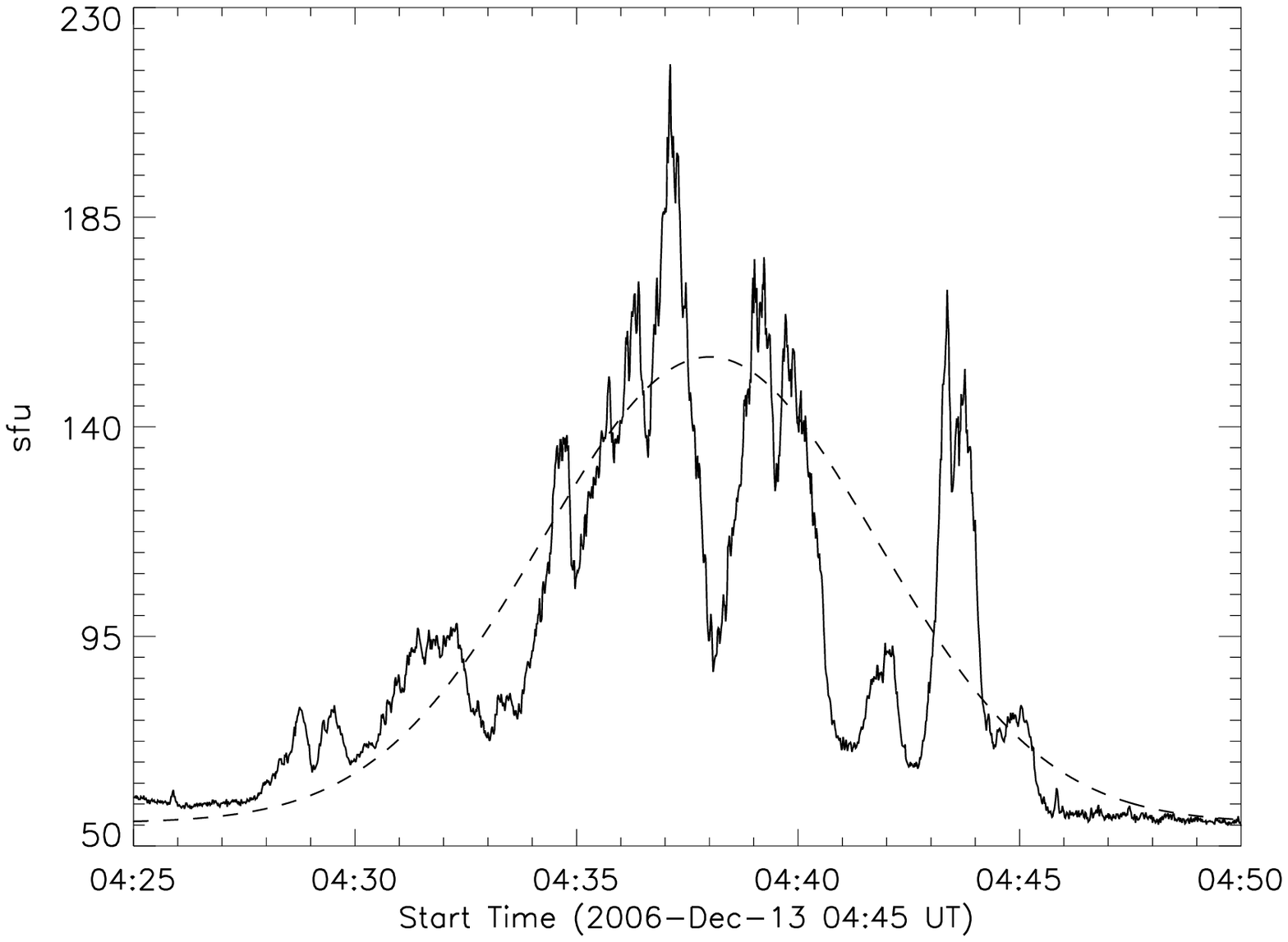}
  \caption{Left panel presents the procedure of QPP extractions. (a) is the observational result, (b) is the smoothed result which filter out the high
  frequency noise components, (c) is the background emission of QPP, and (d) is the pulsating component. According to the profiles,
  the radio bursts can be plotted into 4 paragraphs of QPP which marked as A, B, C, D, and separated by vertical dash-dotted lines, respectively.
  The right panel presents the evolution of VLP paragraph D, here the dashed curve is a Gaussian fitted result.}
\end{center}
\end{figure}

Fig.3 presents an example of the above QPP extraction procedure.
Here, we note that the existence of saturation effects at several
segments may affect the analysis results. Such saturation effects
exist at 02:25 -- 02:25:30, 02:41:30-02:43, 02:52 -- 02:54,
03:01:40 -- 03:05, and 03:32 -- 03:36. Additionally, from the
above method, the pulsating component ($F_{p}$) is a quantity
relative to the background emission ($F_{b}$), and $F_{b}$ is an
average level by smoothing the flux curves ($F$) in a wide sliding
window. From the left panel of Fig.3 we know that $F_{b}$ is much
greater in the flare/CME event than the quiet sun emission (before
the event, i.e. before 02:20 UT), it is possibly mainly associated
with thermal plasmas in the flaring region. So the actual
radiation flux in QPP is always positive. The negative $F_{p}$ is
only a relative result respect to $F_{b}$.

\section{Main Features of QPPs}

Based on the above observation data and analysis methods, we find
that there are 5 classes of QPPs, including VLP, LPP, SPP, slow-
and fast-VSP, coexisting associated with the flare/CME event on 13
December, 2006. They form a broad hierarchy of QPPs with different
timescales. All the QPPs are occurred during the enhancement of
soft X-ray obtained by GOES satellite. From the right panel of
Fig.1 we find that there is no radio burst occurred before 02:20
UT and after 04:50 UT, and the dynamic spectrogram in Fig.2 is
almost uniform pattern during these two time sections. We also use
Fourier analysis to these two sections of observation data, and
find no pulsating evidence. These facts indicate that there is no
detected pulsating structures before and after the flare/CME
event. All the QPPs are associated with the above mentioned
flare/CME event.

\subsection{Very Long-period Pulsation (VLP)}

The most remarkable implication of Fig.1, Fig.2, and Fig.3 may be
the quasi-periodic pulsations with very long-periods. According to
the profile, the radio bursts can be plotted into 4 paragraphs of
QPP which marked as A, B, C, D, and separated by vertical
dash-dotted lines in Fig.3, respectively.

\begin{figure}
\begin{center}
  \includegraphics[width=8.2cm]{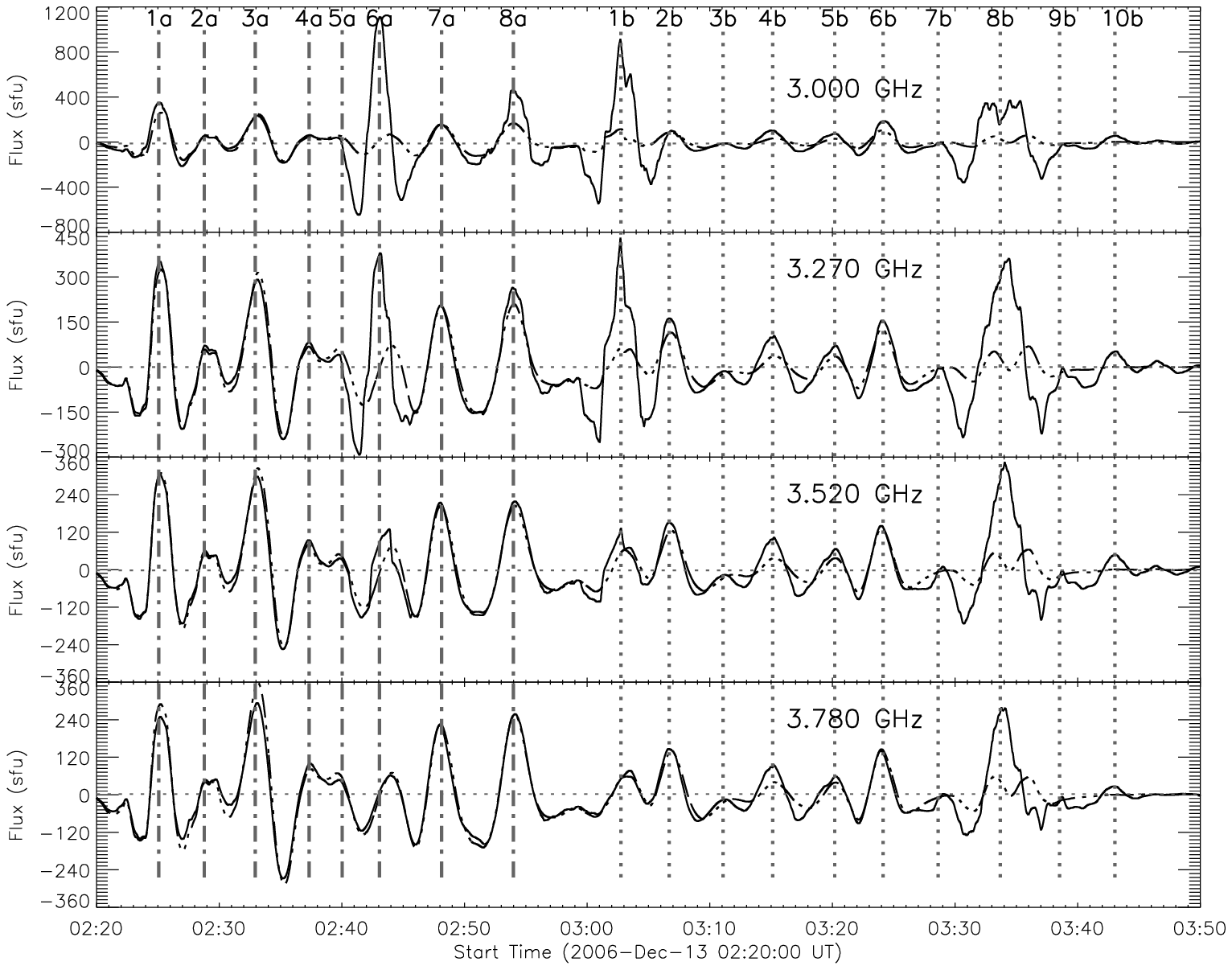}
  \includegraphics[width=8.2cm]{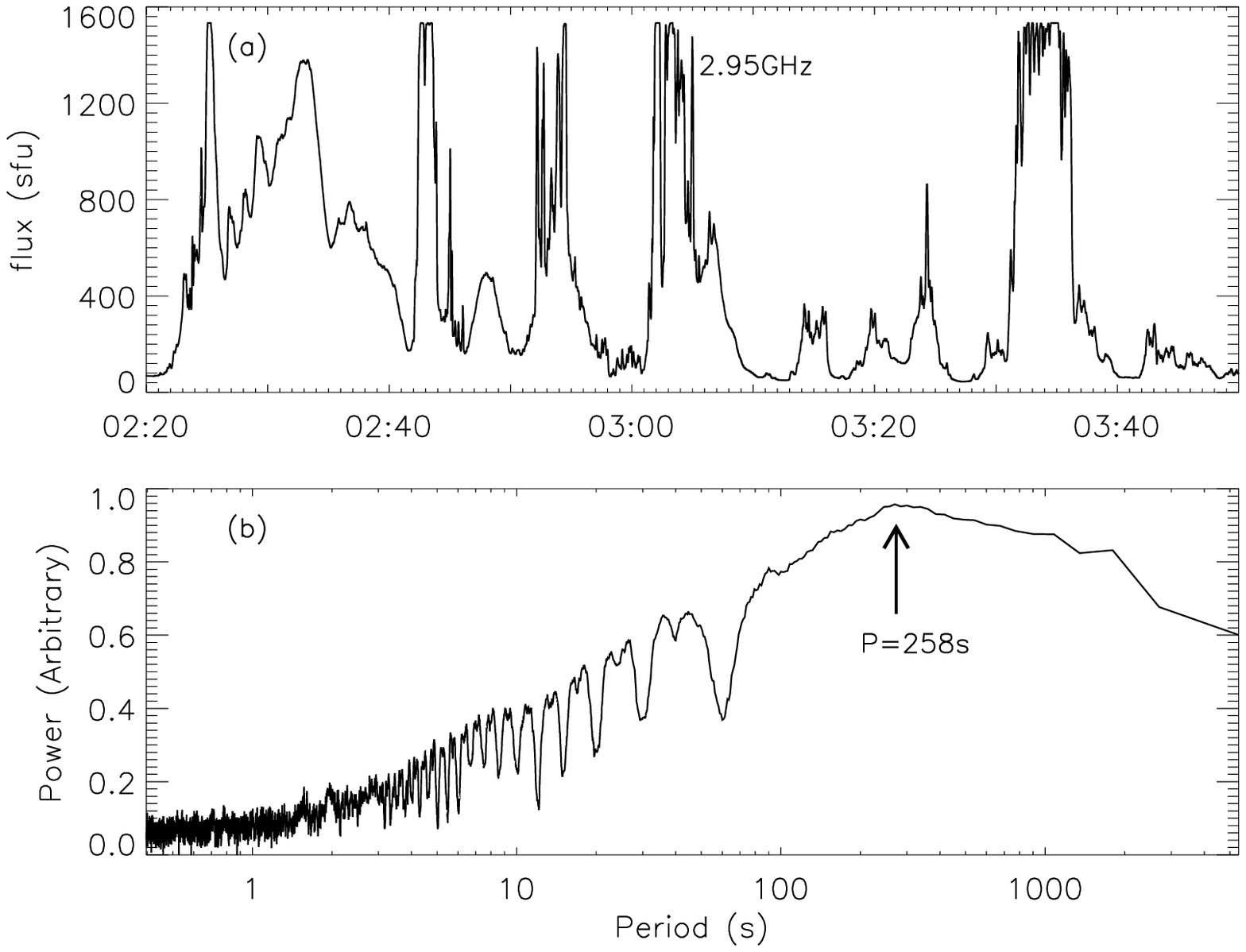}
   \caption{VLP occurred at 02:20 -- 03:50 UT, 13 Dec. 2006 at frequency of 2.60 -- 3.80 GHz observed at SBRS/Huairou.
   Left panel is pulsating components at different frequencies. The solid and dotted curves represent the right and left circular
   polarized component, respectively. Values of pulsating components are relative to the background emission $F_{b}$. In right panel
   (a) is a profile of radio emission at frequency 2.95 GHz. (b) is a result of
   Fourier analysis (FFT) at the corresponding frequency, which indicates the period of the LPP is 258 s.}
\end{center}
\end{figure}

The left panel of Fig.4 is pulsating components at different
frequencies of the VLP including paragraph A and B occurred at
02:20 -- 03:50 UT, 13 Dec. 2006 at frequency of 2.60 -- 3.80 GHz,
which indicates that the VLP is a broadband pulsation and goes
beyond the possible frequency range of the telescope. The solid
and dotted curves represent the right and left circular polarized
component, respectively. Values of pulsating components are
relative to the background emission $F_{b}$. There are 18 pulses
in these paragraphs. The magnitudes of the pulsating emission flux
at each pulse are in the range of 200 -- 1100 sfu at both circular
polarizations. The right panel is a result of Fourier analysis
(FFT) at frequency of 2.95 GHz, the peak indicates that the period
is 258 s.

In fact, there is an obviously gap between pulse 8a and 1b which
divides them into 2 paragraphs:

Paragraph A includes pulses numbered as 1a -- 8a (vertical
dot-dashed lines) with period of 3.1 -- 5.8 minutes. According to
the Fourier analysis we may obtain the period is about 248 s
(about 4.1 minutes). The averaged magnitudes of the pulsating
emission flux is about 350 sfu. However, there are several pulses
(pulse 2a, 3a and 6a), of which values of the pulsating emission
flux are disturbed by saturation effects, and this make the
analysis with some uncertainties that we can not get a reliable
trend from pulse 1a to 8a.

Paragraph B includes pulses numbered 1b -- 10b (vertical dotted
lines) at 03:01 -- 03:45 with period of 4.0 -- 5.1 minutes, and
from the Fourier analysis we may obtain the period is about 270 s
(4.5 minutes). The averaged magnitudes of the pulsating emission
flux is about 300 sfu. The saturation effect is also disturbed
pulses 1b and 8b in same way. However, we can also get the trend
that the magnitudes of the pulses increase from 1b to 8b, and then
decrease from 8b to 10b. Straightforwardly, the periodicity of
paragraph B is more stable than that of paragraph A. At the same
time, the magnitude of the pulsating emission flux at each pulse
in paragraph A is higher than that in paragraph B.

As we mentioned above, the time intervals between two adjacent
pulses on the QPP structure can represent the period of QPP. Fig.5
presents the evolution of these time intervals in VLP paragraph A
(in panel a) and B (in panel b). We may find that the period of
paragraph A increases from about 3 min at the beginning to about
5.8 min at the end, and the period of paragraph B increases slowly
from about 4 min at the beginning to about 5.1 min at the end.
This slowly increasing is consistent with the evolution of MHD
wave mode under coronal conditions at quasi-periodic phase
(Roberts \& Edwin, 1984).

\begin{figure}
\begin{center}
   \includegraphics[width=8.2cm]{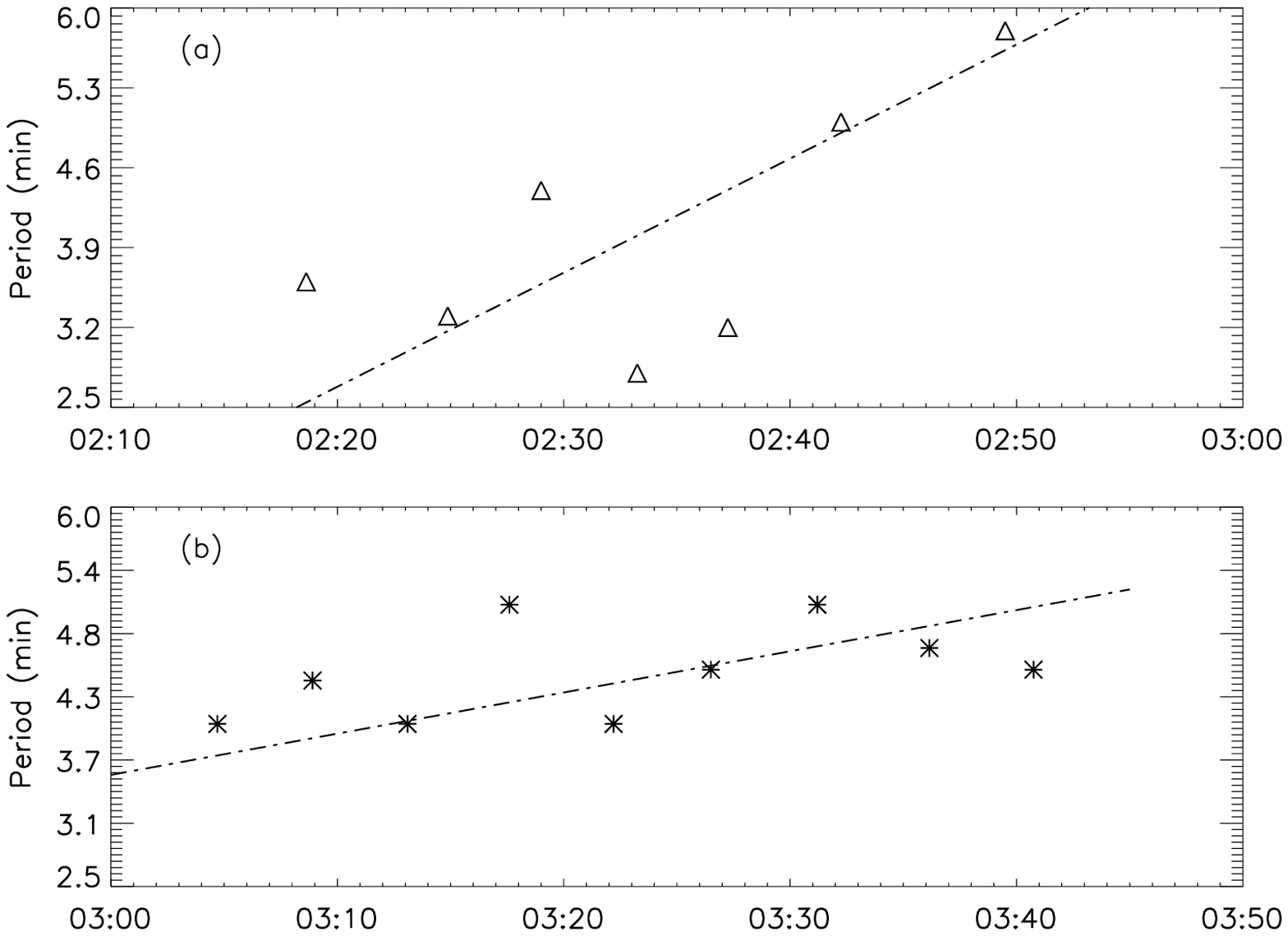}
   \caption{Evolution of the time intervals between adjacent pulses in VLP paragraphs. (a) is in paragraph A, and (b) is in paragraph B.
   the dot-dashed lines are linear squared fitted results.}
\end{center}
\end{figure}

The main difference between the two paragraphs is the evolution of
polarization degree. The left panel of Fig. 4 gives some evidence
that there is a circular polarization reversion from right
circular polarization to left circular polarization around the
frequency of 3.40 GHz in paragraph A. However, in paragraph B,
there is no such revisions, and the polarization degree is keeping
positive. However, because of saturation effects on both paragraph
A and paragraph B, we have no strictly proof to distinguish
paragraph A from paragraph B.

Fig.3 presents two other paragraphs of VLP after 03:50 UT:
paragraph C starts at 03:56 and ends at 04:22 UT, the period is
about 260 s (4.3 min), and the averaged magnitude of the pulsating
emission flux is about 200 sfu; paragraph D starts at 04:28 and
ends at 04:47., the period is about 220 s (3.7 min), and the
averaged magnitude is about 90 sfu, and behaves strongly right
circular polarization. This paragraph has no saturation effect,
the right panel of Fig.3 presents its magnitude evolution, the
dashed curve is a Gaussian fitted result. It shows that the
pulsating process increases gradually, reaches to a maximum, and
then decreases.

As a whole, from paragraph A, B, C to D, the averaged magnitude of
the pulsating emission flux decreases from 350 sfu to 90 sfu,
gradually. The period reaches to the maximum (270 s) at paragraph
B, and then decreases gradually to a minimum (220 s) at paragraph
D.

In addition, there is an obvious frequency drifting at each pulse
in VLP ($R_{spdf}$). Fig.6 presents an example of a method to
calculate the single pulse frequency drifting rate of QPP. The
dot-dashed line denotes the time of the maximum flux intensity at
each frequency, its slope presents the frequency drift rate
($R_{spdf}$) is about -175 MHz/s. Using this method, we obtained
$R_{spdf}$ at each pulse of paragraph A as 21, 47, -183, -440,
-215, 55, -175, 50 MHz/s, and at each pulses of paragraph B as 25,
48, 65, 389, 46, -50, -76, 313 MHz/s, respectively. In addition,
we can also find that $R_{spdf}$ begins from a positive value,
then decreases to negative value, and ends with another positive
value in each paragraph. Using the same method, single pulse
frequency drifting rates of paragraph C and D are in the range of
-50 -- $\sim$ 190 MHz/s and -135 -- $\sim$ 265 MHz/s,
respectively. The main properties of VLPs are summarized in Table
1.

\begin{figure}
\begin{center}
  \includegraphics[width=8.2cm]{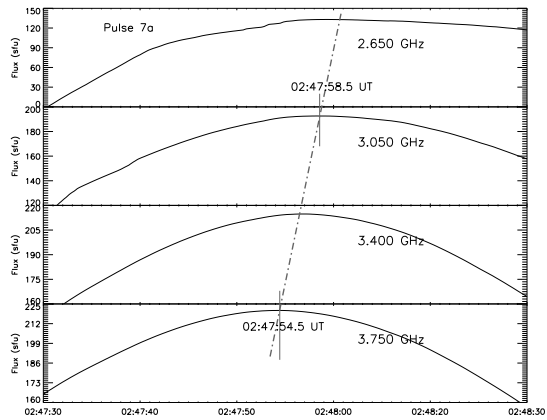}
   \caption{Drifting characteristics of a very long period pulsation (here is an example from the pulse 7a marked in the left panel of Fig 4),
   the dot-dashed line indicates positions of the maximum flux intensity at each curve, and its slope indicates the
   frequency drift rate at the corresponding pulse.}
\end{center}
\end{figure}

\begin{table}\def~{\hphantom{0}}
  \begin{center}
  \caption{List of VLP, LPP, and SPP occurred in the flare event on 2006-12-13. $f_{0}$, D, P, r, $R_{spdf}$, and $b_{w}$ are central frequency, duration,
  average period, polarization degree, average single pulse frequency drifting rate, and frequency bandwidth, respectively.}
  \label{tab:kd}
  \begin{tabular}{lccccccccccccc}\hline
Class  & Time (UT)         & $f_{0}(MHz)$ & D(s) &   P(s) &  r($\%$)       & $R_{spdf}$(MHz/s)& $b_{w}$(MHz) \\\hline
VLP    & 02:23--02:56      &  3200        & 1980 &   248  & $ 8 \sim  10$  & -440$\sim$ 55    &  $>$1200     \\
       & 03:01--03:45      &  3200        & 2640 &   270  & $ 8 \sim  12$  & -76 $\sim$389    &  $>$1200     \\
       & 03:56--04:22      &  3200        & 1560 &   260  & $ 15 \sim 35$  & -50 $\sim$190    &  $>$1200     \\
       & 04:28--04:47      &  3200        & 1140 &   220  & $ 25 \sim 40$  & -135 $\sim$265    &  $>$1200     \\\hline
LPP    & 02:26--02:34      &  3200        &  480 &   70   & $ 10 \sim 20$  &    -610          &   $>$1200    \\
       & 02:38:50--02:40:00&  2800        &  70  &   16   & $ 10 \sim 20$  &    -5.8          &    300       \\
       & 02:52--02:55      &  3200        &  180 &   35   & $ 45 \sim 65$  &    -205          &   $>$1200    \\
       & 03:23:35--03:24:50&  3200        &  75  &   15   & $ 55 \sim 75$  &     78.5         &   $>$1200    \\\hline
SPP    & 02:27:50--02:28:20&  2850        &  30  &   8.0  & $ 65 \sim 85$  & -1500$\sim$2500  &   400        \\
       & 02:59:20--03:00:20&  3050        &  60  &   7.5  & $ 80 \sim 95$  & -4000$\sim$3500  &   750        \\
       & 02:24:35--02:24:47&  2800        &  12  &   1.2  & $ 85 \sim 100$ &  $\sim$5000      &   350        \\\hline
  \end{tabular}
 \end{center}
\end{table}

Comparing VLP paragraph A with the Fig.2 of Minoshima et al
(2009), we find that its flux profile is very similar to the
microwave light curves of 9.4 GHz, 17 GHz, and 34 GHz taken with
Nobeyama Radio Polarimeter (NoRP) and the hard X-ray light curves
in 25-40 keV, 40-60 keV, and 60-100 keV energy band taken with
Reuven Ramaty High Energy Spectroscopic Imager (RHESSI). These
facts may indicate that pulsating microwave emissions is most
possible to be associated strongly with the non-thermal energetic
electrons.

\subsection{Long-period Pulsation (LPP)}

During the flare/CME event, we totally distinguish 4 segments of
LPP listed in Table 1. Fig.7 shows an example of LPP occurred at
02:26 -- 02:40 UT, 13 Dec. 2006 observed at SBRS/Huairou. The left
panel of Fig.7 is the dynamic spectrogram which indicates the
period (the time interval between adjacent pulse stripes) is in
the range of 55-95 seconds, with the average of about 70 seconds.
In right panel of Fig.7, (a) is the profile of radio emission at
frequency 2.84 GHz, the dashed curve is a Gaussian fitted result
which shows the magnitude evolution of increasing to a peak and
then decreasing, (b) is a result of Fourier analysis (FFT) at the
corresponding frequency, which indicates the period of the LPP is
70 s, agreeing with the dynamic spectrogram. The magnitude of the
pulsating emission flux at each pulse is in the range of 70 -- 220
sfu at both circular polarization components with the polarization
degree of about $10\sim20\%$. And the emission bandwidth is also
beyond the two borders of the frequency range of the telescope of
2.60 -- 3.80 GHz, which is similar to that of VLP.

The other LPPs occurred at 02:38:50 -- 02:40:00 with mean period
of 16 s (and this LPP has some superfine structures, we will
discuss them in the following sections, see Fig.11), at 02:52 --
02:55 UT with mean period of 35 s, and at 03:23:35 -- 03:24:50
with mean period of 15 s. Q-factors of these four LPPs are around
5-7 which is approximately a constant.

\begin{figure}
\begin{center}
  \includegraphics[width=8.2cm, height=8.5cm]{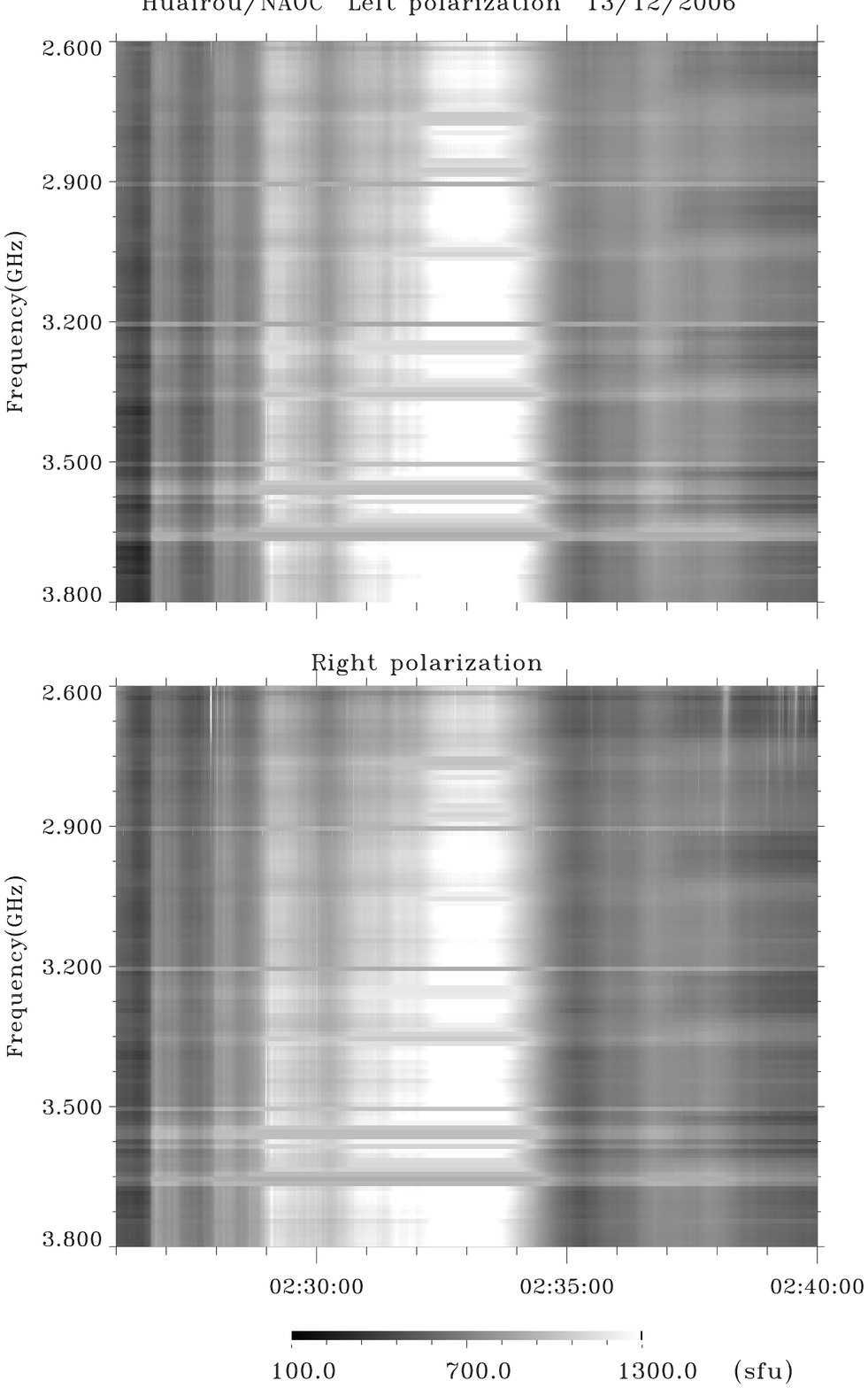}
  \includegraphics[width=8.cm, height=8.cm]{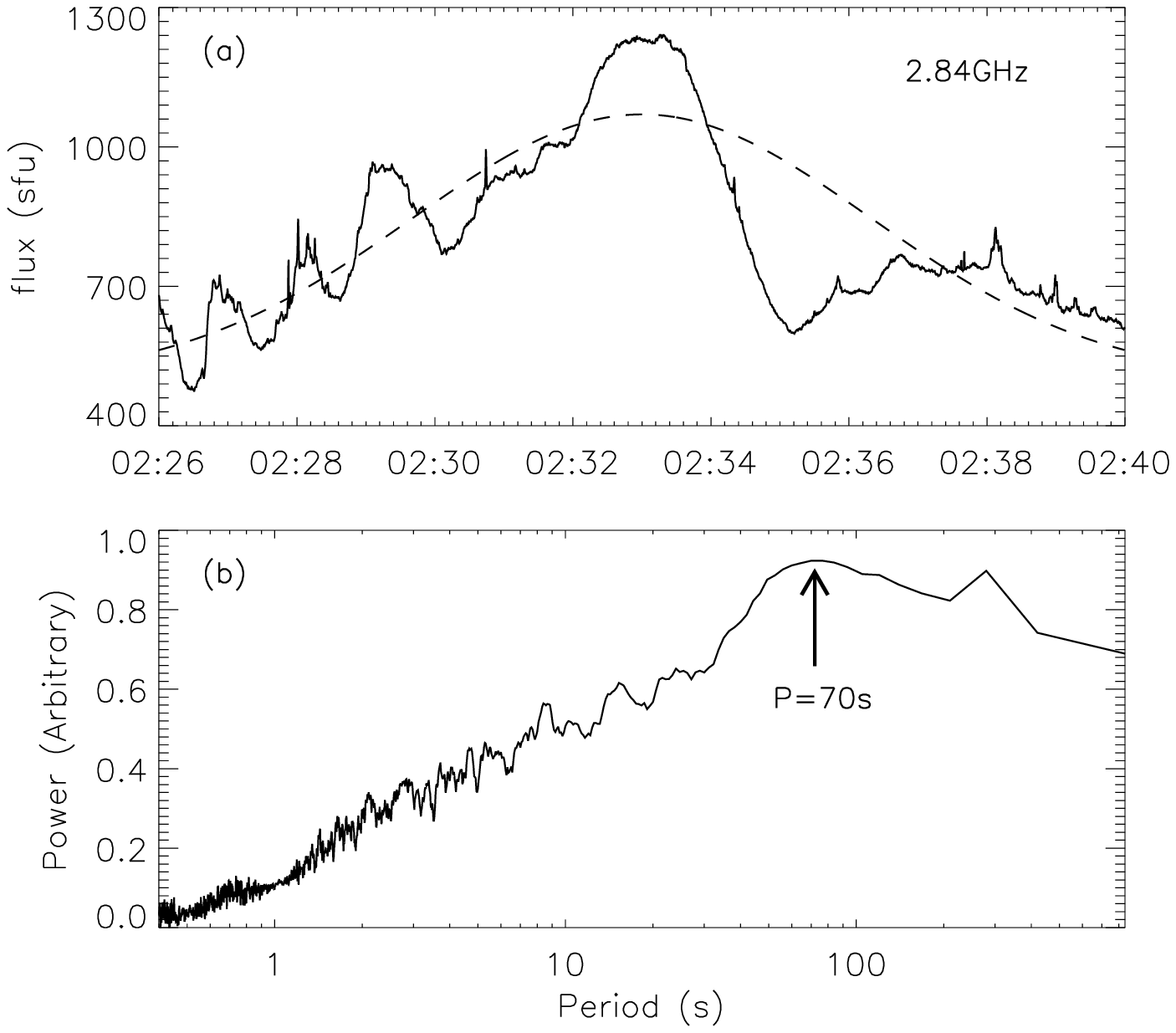}
  \caption{The left panel is a spectrogram of a LPP. In right panel,
  (a) is the profile of radio emission at frequency 2.84 GHz, the dashed curve is a Gaussian fitted result.
  (b) is the result of Fourier analysis (FFT) at the corresponding frequency, which indicates the period of the LPP is 70 s. }
\end{center}
\end{figure}

Using the same method presented in Fig.6, we also find that single
pulse frequency drift rate exists in LPP. The average value in
these 4 segments are -610 MHz/s, -5.8 MHz/s, -205 MHz/s, and 78.5
MHz/s, respectively. And these values are in the same order as
VLP. From the comparison between VLP and LPP we find that there
are several points of similarity: analogous frequency bandwidth,
single pulse frequency drift rates, polarization degree, and
Q-factors. These similarities imply that VLP and LPP should be
associated originally with each other.

\subsection{Short-period Pulsation (SPP)}

During the flare/CME event, we identified 3 segments of SPP. They
are listed also in Table 1. Fig.8 presents an example of SPP
occurred at 02:59:20 -- 03:00:20 UT, 13 Dec. 2006. The left panel
of Fig.8 is the dynamic spectrogram which indicates the period
(the time interval between adjacent pulse stripes) is in the range
of 4.7 -- 7.2 s, with the average of 6.2 s. In right panel of
Fig.8, (a) is the profile of radio emission at frequency 2.95 GHz,
the dashed curve is a Gaussian fitted result which shows the
magnitude evolution which also behaves as increasing to a peak and
then decreasing, (b) is a result of Fourier analysis (FFT) at the
corresponding frequency, which indicates the period of the SPP is
6 s, consistent with the result of dynamic spectrogram. The
emission frequency bandwidth is about 600 -- 900 MHz, which is
narrower than that of VLP and LPP. The magnitude of pulsating
emission flux at each pulse is in the range of 40 -- 115 sfu with
right circular polarization, while there is almost no pulsating
component in the left circular polarization. From the dynamic
spectrogram, we may also find that SPP has a strong right circular
polarization. Additionally, there are obviously frequency drifts
at each pulse of the SPP, its value is in the range of -4.0 GHz/s
to 3.5 GHz/s. And this frequency drift rate is very close to the
drifts of type III bursts (Ma et al, 2006). However, the work of
Aschwanden \& Benz (1986) indicates that the frequency drift rates
of QPPs are considerably different from type III bursts in the
similar frequency range, and these two kinds of microwave fine
structures are intrinsically different from each other.

\begin{figure}
\begin{center}
  \includegraphics[width=8.cm, height=8.cm]{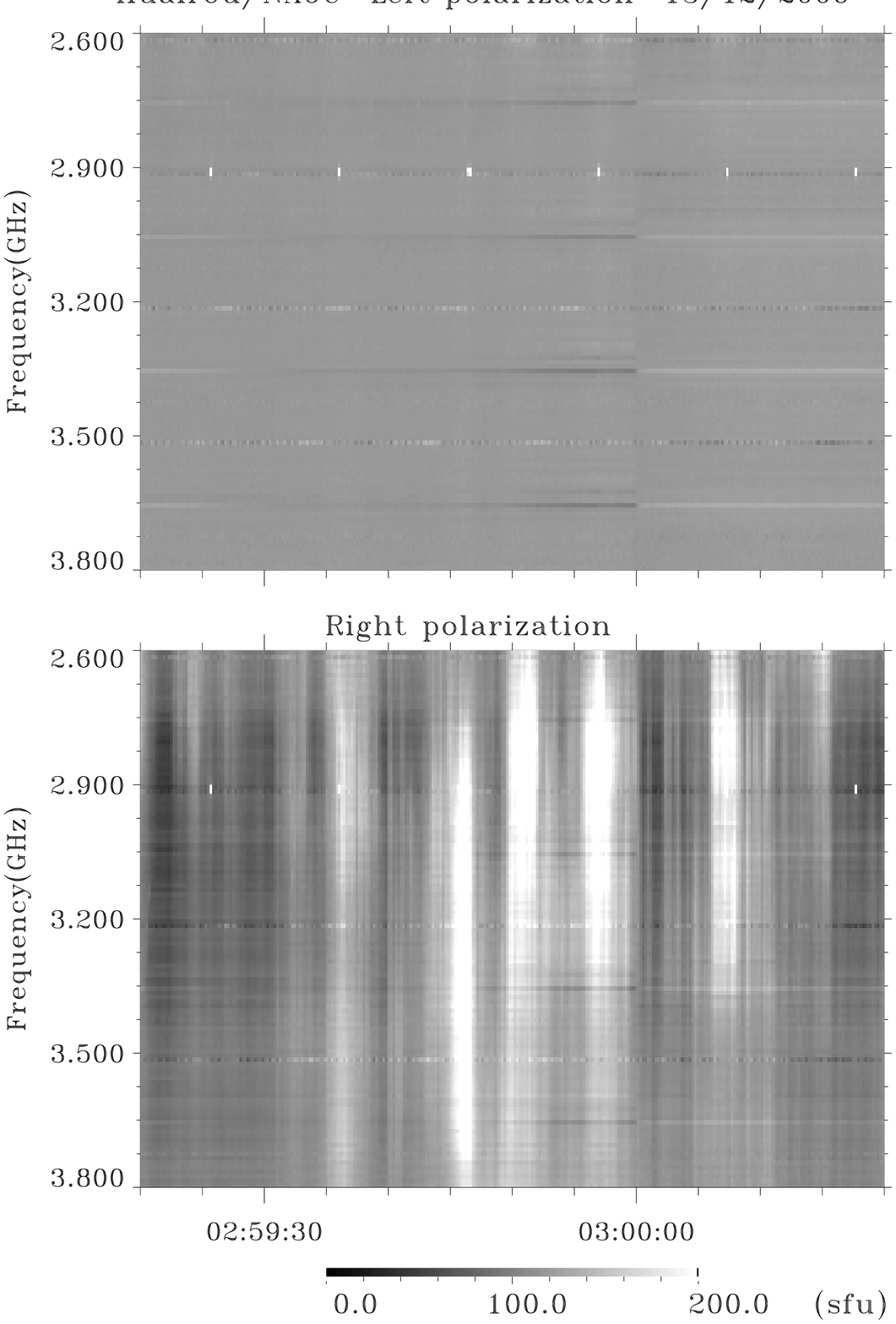}
  \includegraphics[width=8.cm, height=7.5cm]{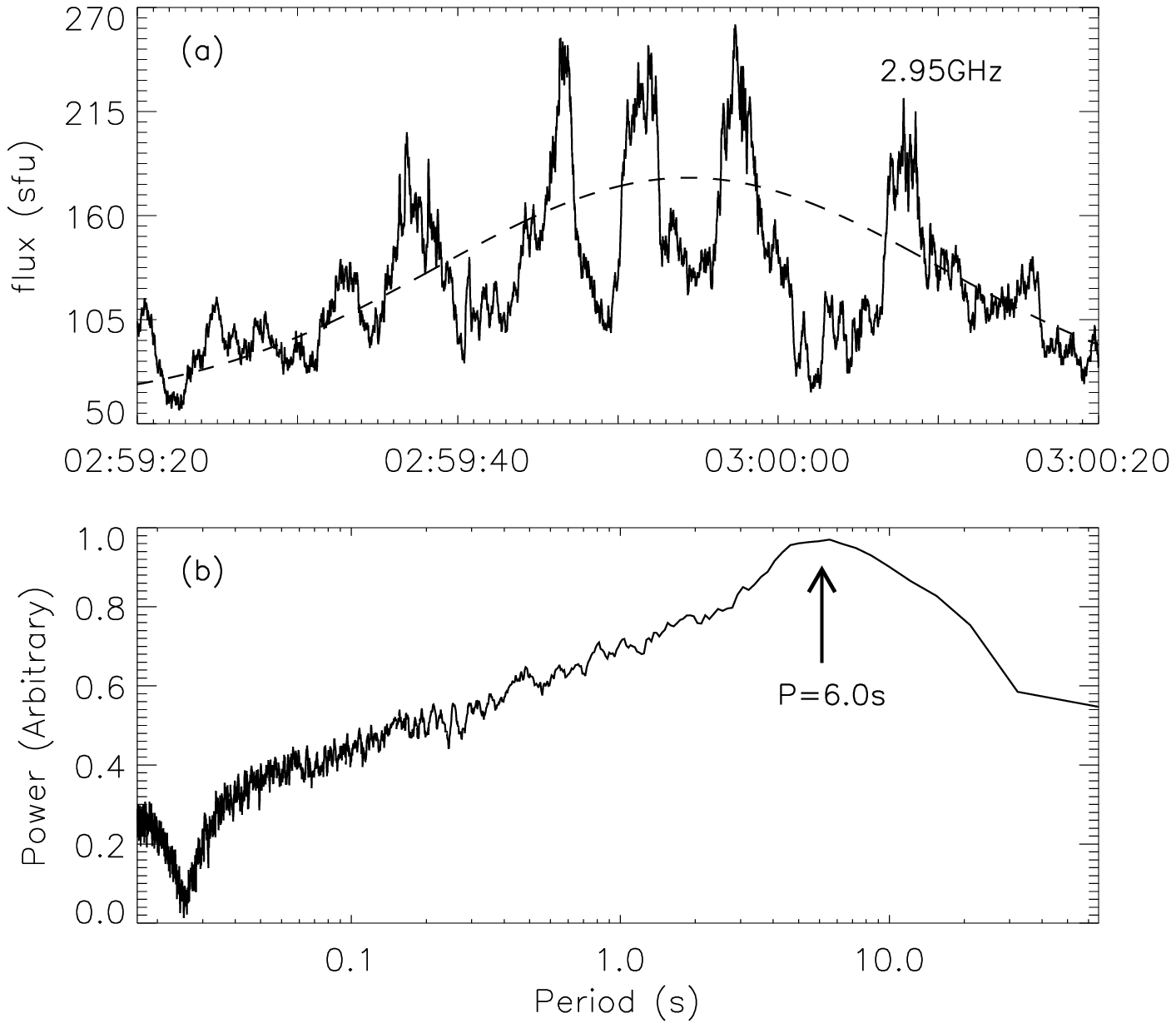}
  \caption{Left panel is a spectrogram of a SPP example. In right panel,
  (a) is the profile of radio emission at frequency 2.84 GHz,  the dashed curve is a Gaussian fitted result.
  (b) is the result of Fourier analysis (FFT) at the corresponding frequency, which indicates the period of the SPP is 6 s.}
\end{center}
\end{figure}

In fact, from the dynamic spectrogram of Fig.7 we may find that
there is a SPP superposed on the LPP at 02:27:50 -- 02:28:20 UT in
the frequency range of 2.60 -- 3.00 GHz with bandwidth 400 MHz,
its period is about 8.0 seconds, and the polarization degree is
about 65 -- 85\%. And the magnitude of the pulsating emission flux
at each pulse is in the range of 50 -- 120 sfu. The single pulse
frequency drift rate is about from -1.5 GHz/s to 2.5 GHz/s.
Another SPP occurred at 02:24:35 -- 02:24:47 UT with the averaged
period of about 1.2 seconds at frequency of 2.60 -- 2.95 GHz, the
bandwidth is about 350 MHz, and the single pulse frequency drift
rate is about 5.0 GHz/s. From these facts we may find that SPP is
completely different from VLP and LPP with parameters of
polarization degree, frequency drift rate, and frequency
bandwidth, etc. These differences imply that SPP may have a
different generation mechanism from VLP or LPP.

\subsection{Very Short-period Pulsation (VSP)}

More than 40 cases of VSP with period $P<1.0$ s in the frequency
range of 2.60 -- 3.80 GHz associated with the flare/CME event on
13 December 2006 were reported in the work of Tan et al (2007) and
Tan (2008). These VSPs are quasi-periodic, broad bandwidth, and
ubiquitous in all phases of the flare/CME event. Among these VSPs,
the right circular polarization is very strong. Table 2 of Tan
(2008) lists all their observable features. From that list we may
classify them into 2 sub-classes:

1) slow-VSP, the period is in deci-second, $0.1s<P<1.0s$. The left
panel of Fig.9 is an example of slow-VSP occurred at 03:25:40 --
03:25:47 UT, 13 Dec. 2006 at frequency of 2.60 -- 2.90 GHz.
Totally there are 10 slow-VSP cases distinguished from the
observations, which are associated with the flare event. And all
of them occurred after the flare peak. The period of slow-VSP is
in the range of 110 -- 416 ms. The duration of each slow-VSP is in
the range of 2.2 -- 23 s, the single pulse frequency drift rate is
in the range of 6.5 -- 20 GHz/s and is always positive, the
emission frequency bandwidth is in 250 -- 700 MHz. The
polarization is very strong, similar to that of SPP. The magnitude
of the pulsating emission flux is in the range of 20 -- 95 sfu. We
may also obtain the global frequency drift rate of slow-VSP, and
the value is from -31.4 MHz/s to 32.5 MHz/s.

The left panel of Fig.9 shows that the magnitude of pulses
increases gradually, and the period decreases from 480 ms to 310
ms slowly. However, the sepctrogram becomes continuum after
03:25:47 UT. The possible reason is that it is mixed by some other
emissions, and the left panel of Fig.9 is only a partial section
of the slow-VSP. By scrutinizing other slow-VSPs, we find that
some of them have increasing magnitudes and decreasing periods,
and the others have decreasing magnitudes and increasing periods.

\begin{figure}
\begin{center}
  \includegraphics[width=7.5cm, height=8.0cm]{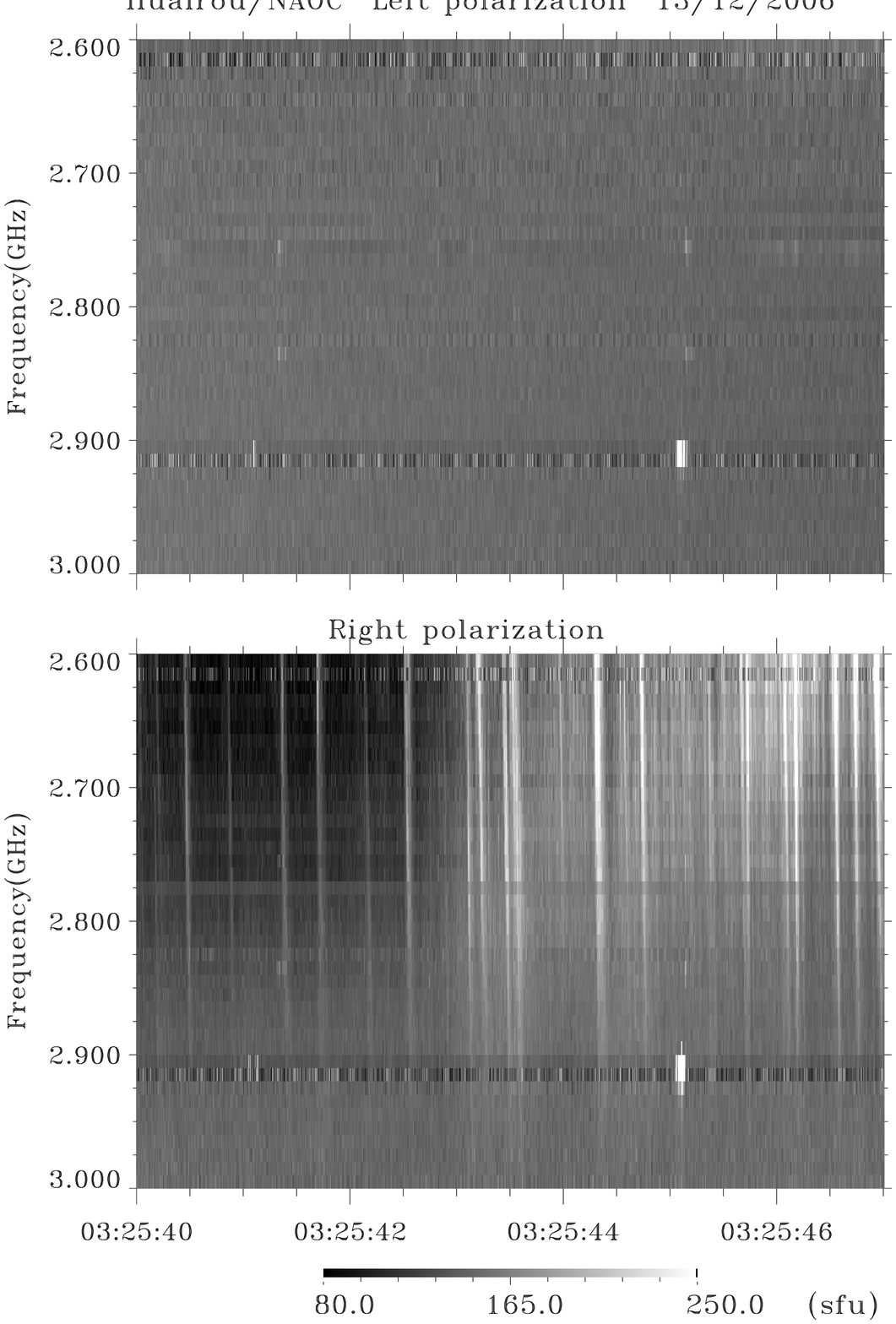}
  \includegraphics[width=7.5cm, height=8.0cm]{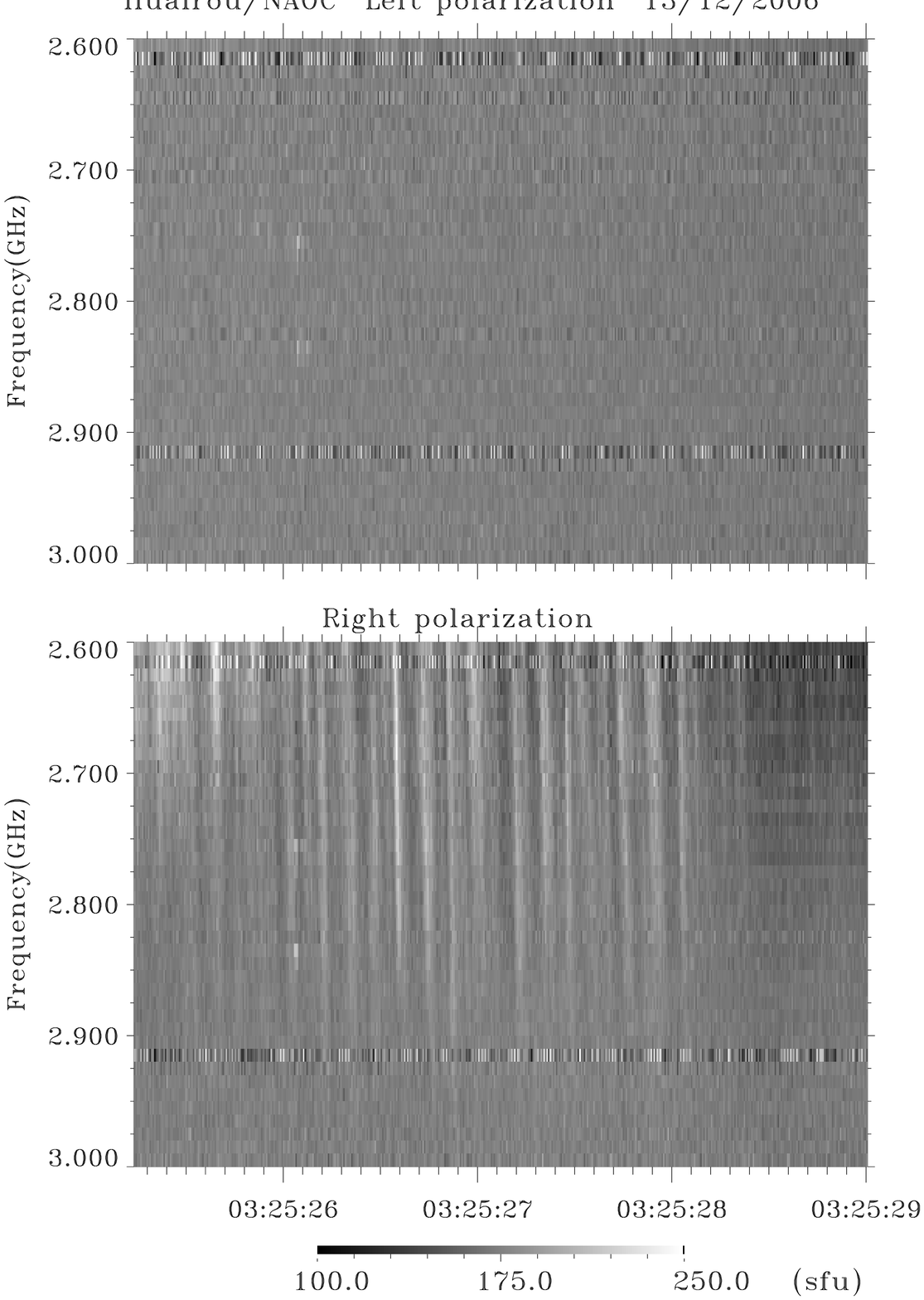}
  \caption{Examples of a slow-VSP (extraordinary VSP), and a fast-VSP (ordinary VSP). In each panel,
  the upper and the lower part is the left- and right- circular polarization component, respectively.}
\end{center}
\end{figure}

Ji$\check{r}$i$\check{c}$ka \& Karlick$\acute{y}$ (2008) also
found a similar slow-VSP with the period of about 150 ms and the
single pulse drift rate is about -17 GHz/s associated with another
solar flare event.

2) fast-VSP, the period is in centi-second, $P<0.1s$. The right
panel of Fig.9 is an example of fast-VSP occurred at 03:25:25 --
03:25:29 UT, 13 Dec. 2006 at frequency of 2.60 -- 2.90 GHz. There
are more than 30 fast-VSPs distinguished from observations
associated with the flare event, and they occurred in all phases
of the flare (including the rising, peak, vale, and decaying
phases). The period of fast-VSP is in the range of 31 -- 90 ms.
The duration of each fast-VSP is in the range of 2.3 -- 16 s, the
single pulse frequency drift rate is from 8.33 GHz/s to $>$30
GHz/s, the emission bandwidth is in 220 -- 400 MHz. The
polarization degree is slightly higher than that of slow-VSP. The
magnitude of the pulsating emission flux is in the range of 15 --
70 sfu, slightly less than that of slow-VSP. The global frequency
drift rate of the pulsating event is from -86.2 MHz/s to 34.6
MHz/s, which is very analogous to that of slow-VSP.

The right panel of Fig.9 presents that the magnitude of pulses
increases slowly from 160 sfu to a peak of 264 sfu, and then
decreases to 145 sfu, and the period increases slowly from 75 ms
to 104 ms. Statistically analyzing other fast-VSP, we find that
there are 35\% fast-VSPs have the above evolutions, and 37\%
fast-VSPs have evolution of increasing magnitudes and decreasing
periods, and 28\% fast-VSPs have evolution of decreasing
magnitudes and increasing periods. This implies that most of the
fast-VSPs are possibly also partially sectional.

Let us come back to the Table 2 in the paper of Tan (2008), we may
find that emission frequency bandwidths of the most VSPs are
narrower than that of VLP and LPP. Additionally, from the
characteristics of the flux profiles, we find that there is
another interesting feature: the VSPs can be classified into two
classes by another criteria:

1) Ordinary VSP, the width of pulses at half maximum ($W_{p}$) is
approximated to a half of the gap between two adjacent pulses
($W_{g}$): $W_{p}\approx 0.5W_{g}$. Most of the fast-VSPs belong
to ordinary VSP (see the right panel of Fig.9). In a standard
sinusoidal curves: $W_{p}= \frac{1}{2}W_{g}$. The ordinary VSPs
are sinusoidal-like pulsations.

2) Extraordinary VSP, $W_{p}\ll W_{g}$. The left panel of Fig.9
shows an example of extraordinary VSP, which period is about 416
ms, $W_{p}$ is about 35 -- 60 ms, and $W_{g}$ is about 365 --380
ms.

Among 40 VSPs associated with the flare/CME event, there are only
3 extraordinary VSPs, which occurred at 03:25:40 -- 03:25:47 UT,
03:28:10 -- 03:28:15 UT, and 03:44:09 -- 03:44:14 UT, and their
periods are 416 ms, 200 ms, and 160 ms, respectively. All
extraordinary VSPs belong to slow-VSP, strongly right circular
polarization, and the magnitude of the pulsating emission flux is
in the range of 30 -- 70 sfu.

In fact, all of the QPPs can be also classified into ordinary QPP
and extraordinary QPP. When we re-scrutinize the scales of $W_{p}$
and $W_{g}$ in other QPPs, we may find that almost all the VLP,
LPP, and SPP are belonging to ordinary QPP.

\subsection{Relationships among QPPs in Multi-timescales}

Fig.10 presents a synthetical comparison of temporal relationships
of the 5 classes of QPPs with different timescales, GOES soft
x-ray flux, and the transport rate of magnetic helicity (dH/dt).
We find that all LPPs and SPPs occurred near peaks of VLP.
Additionally, from the flare impulsive phase to its decay phase,
the duration of LPP is gradually decreased. However, VSPs
distribute in all phases of the flare (rising phase, peak, decay
phase, or vale). And this may imply that VSP is possibly
independent from other classes of QPPs. It seems to be a small
additional component superposed on other classes of QPPs with
longer timescales.

\begin{figure}
\begin{center}
  \includegraphics[width=8.5cm]{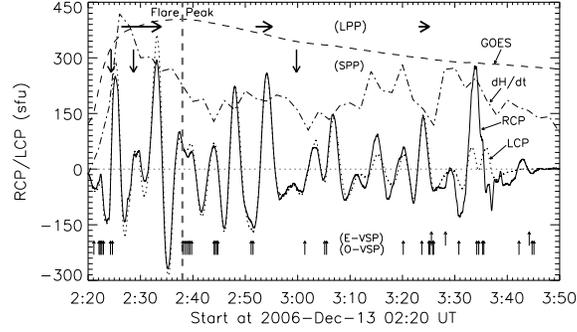}
   \caption{A synthetical comparisons of the temporal relationships among the 5 classes of QPPs, GOES soft x-ray flux (GOES),
   and the transport rate of magnetic helicity (dH/dt). Profiles of the left- and right- circular polarization (LCP, RCP)
   represent the VLP, while the big, moderate and vertical small arrows represent LPP, SPP, and VSP, respectively. The length of the big arrow show
   the duration of LPP. Here O-VSP indicates the ordinary VSP, and E-VSP indicates the extraordinary VSP. }
\end{center}
\end{figure}

The period ratios between different QPPs may be meaningful. Around
the first LPP shown in Fig.10, there occurred 3 classes of QPPs:
VLP (paragraph A), LPP and SPP. Their averaged periods are 4.13
min, 70 s, and 8 s, respectively. Their ratios are 31.0 : 8.75 :
1. The second LPP is just occurred in the gap between paragraph A
and B of VLP, and during this gap we did not distinguish other
kind of QPP. During the third LPP, there occurred VLP (paragraph
A, $P=4.5$ min), LPP ($P=15$ s) and VSP ($P=90\sim110$ ms). The
period ratio is about 2700 : 150 : 1, which is entirely different
from the former ratios. Fig.11 shows an example of concurrence of
two different classes of QPPs occurred at 02:38:50 -- 02:40:00 UT.
The higher hierarchy of QPP is a LPP with period of 16 s, and the
lower hierarchy of QPP is a fast-VSP with period of 77 ms. From
this figure we find that the LPP is fine-structured with a train
of fast-VSPs, and the period ratio is more than 200. And in the
train of fast-VSPs, we find that there is obviously global
frequency drifting rate with value of about 2.5 MHz/s, and this
drift rate is very slow comparing with the single pulse frequency
drifting rates. These evidences imply that there is no obvious
regulations among the period ratios and the weakly links between
the different classes of QPPs.

Additionally, from careful investigation of the microwave
observations there is no evidence of any LPP, SPP and VSP found
around the time intervals of VLP paragraph C and D. It seems that
all the LPP, SPP, and VSP take place around the time intervals of
VLP paragraph A and B.

\begin{figure}
\begin{center}
  \includegraphics[width=7.2cm, height=7.8cm]{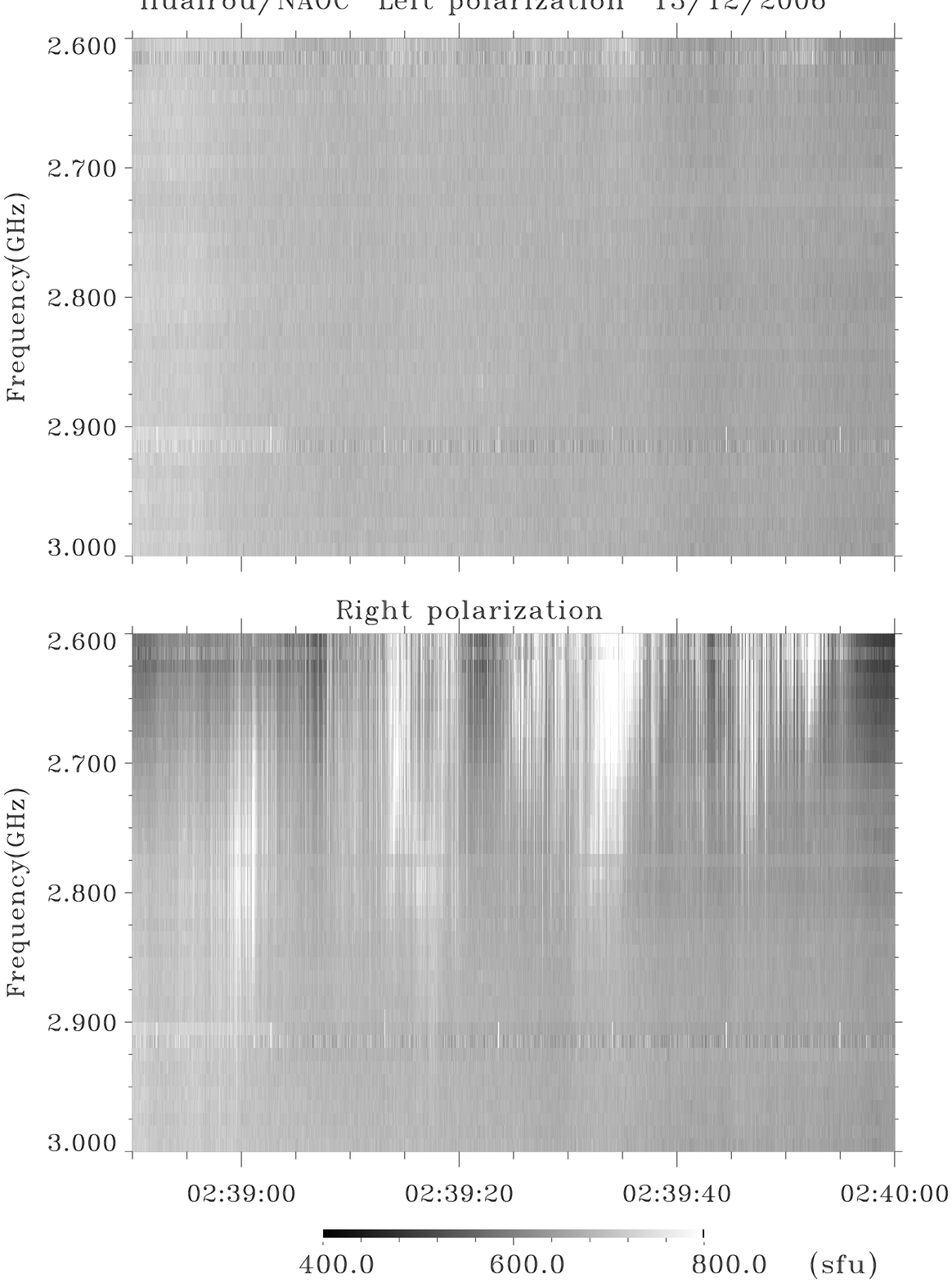}
  \includegraphics[width=8.0cm, height=7.5cm]{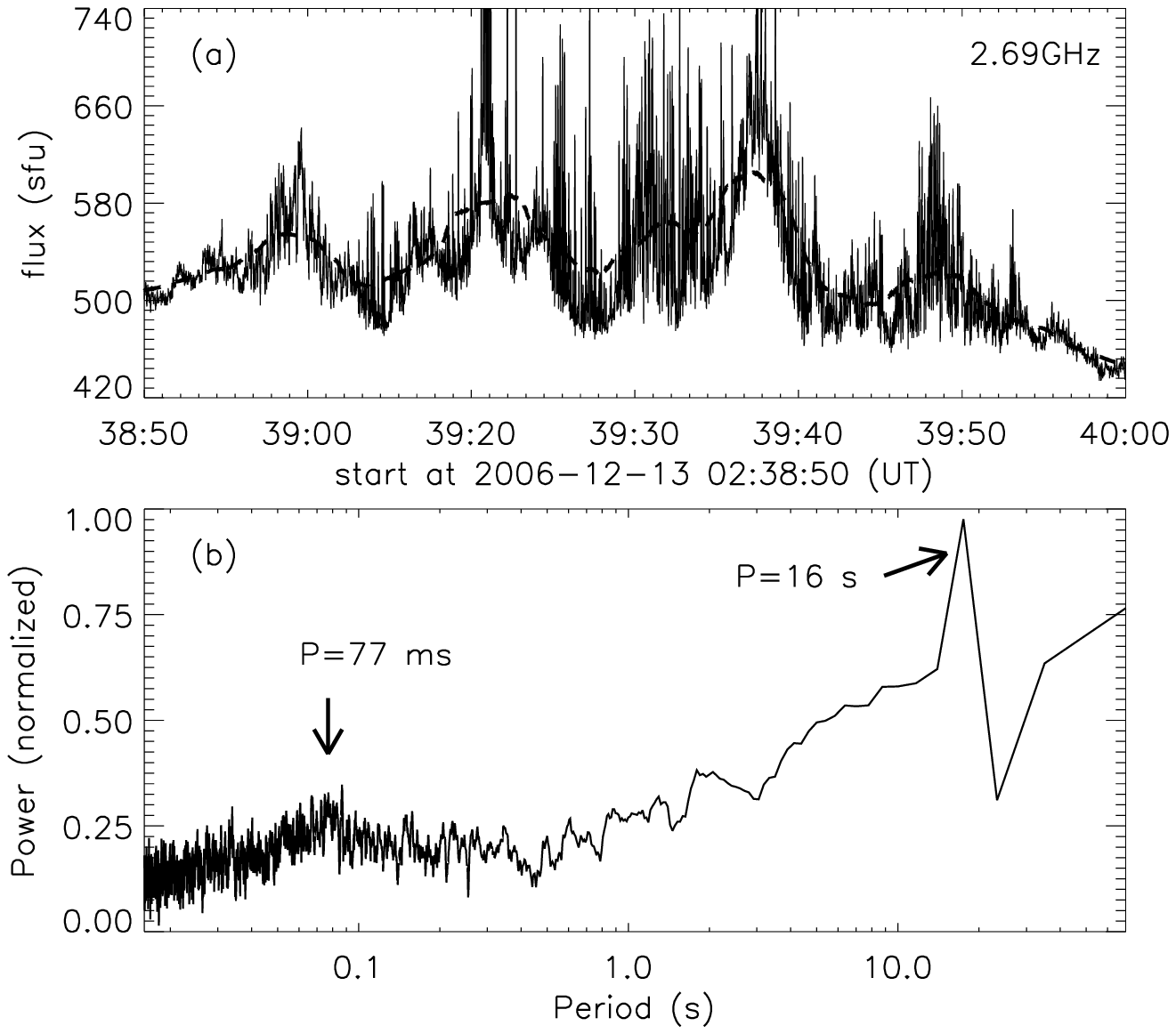}
   \caption{An example of concurrence of two different hierarchies of QPPs. The left panel is the spectrograms of left- and
   right-circular polarization. Panel (a) is a flux curve at frequency of 2.69 GHz, the solid curve indicates a fast-VSP, and the
   thick-dashed curve indicates a LPP. Panel (b) is a result of Fast Fourier Transformation at frequency of 2.69 GHz, the two pesks represent
   a VSP of period 77 ms and a LPP of period 16 s, respectively.}
\end{center}
\end{figure}

Usually, the microwave burst can be regarded as a prompt signal of
non-thermal energetic particles originating from the magnetic
reconnection. So it is meaningful to quest for the relationships
between the magnetic field and QPPs. Here, we introduce the
transport rate of magnetic helicity ($dH/dt$) to describe the
magnetic field behavior. It can be expressed as (Berger \& Field,
1984):

$\frac{dH}{dt}=\oint2(\textbf{\textit{B}}\cdot\textbf{\textit{A}}_{\textrm{p}})\upsilon_{\textrm{z}}dS-\oint2(\mbox{\boldmath
$\upsilon$}\cdot\textbf{\textit{A}}_{\textrm{p}})B_{\textrm{z}}dS$.

Here, $\textbf{\textit{A}}_{\textrm{p}}$ is magnetic vector
potential, ${\upsilon}$ is the fluid velocity, and
$B_{\textrm{z}}$ is the normal component of the magnetic field. In
an open volume, the magnetic helicity may change by the passage of
helical magnetic field lines through the surface (the first term)
and by shuffling horizontal motion of field lines on the surface
(the second term). In an isolated active region $dH/dt$ is a
nonpotential parameter which indicates the dynamic evolution of
the magnetic field mainly in the flaring region (Berger \& Field,
1984). As the flare/CME event occurred in an isolated active
region of AR 10930, we may estimate the quantity for the event by
calculating from the above expression. The work of Zhang et al
(2008) presents that the temporal profile of $dH/dt$ is consistent
with that of microwave burst in the flare/CME event on Dec.13,
2006. So we superposed the curve of $dH/dt$ in the same duration
in Fig.10 (dashed-dotted curve). Here, the cadence of dH/dt is 2
minutes, which is calculated from the observations of the Solar
Optical Telescope on board Hinode (SOT/Hinode, Tsuneta et al,
2008; Kosugi et al, 2007). From the comparison of $dH/dt$ and
QPPs, we find that when QPPs take place $dH/dt$ is positive. This
fact indicates a continuous injection of magnetic helicity. This
may lead to accumulation of nonpotential energy and the magnetic
reconnections in the flaring active region. There are some
variations in the profile of $dH/dt$ during the flaring event.
However, as the cadence of $dH/dt$ is only 2 minutes, we could not
confirm any pulsating features in $dH/dt$.

Actually, we find that VLP, LPP and SPP have much of similarities,
such as broad emission bandwidth, almost at the same level of
Q-factor ($\leq10$), weakly circular polarization, and relatively
slow frequency drift rates, except that they have different
durations, periods, and different magnitude of pulsating emission
flux. They belong possibly to a same class of QPP in a sense. In
order to confirm such viewpoint, we plot all the QPPs in a
logarithmic period-duration space in Fig.12. We find that all of
the VLPs, LPPs, SPPs, and part of slow-VSPs are distributed around
a line. However, almost all fast-VSPs and most part of the
slow-VSPs are distributed far away from the above line
dispersively. This fact implies that fast-VSPs and part of
slow-VSPs may be different originally from VLPs, LPPs, and SPPs.
Hereby, we may classify all QPPs into two groups: group I includes
VLPs, LPPs, SPPs, and part of slow-VSPs of which the longer the
duration corresponds to the longer period; group II includes
fast-VSPs and most part of slow-VSPs of which the period is
dispersive respect to the duration.

\begin{figure}                         
\begin{center}
  \includegraphics[width=8.5cm]{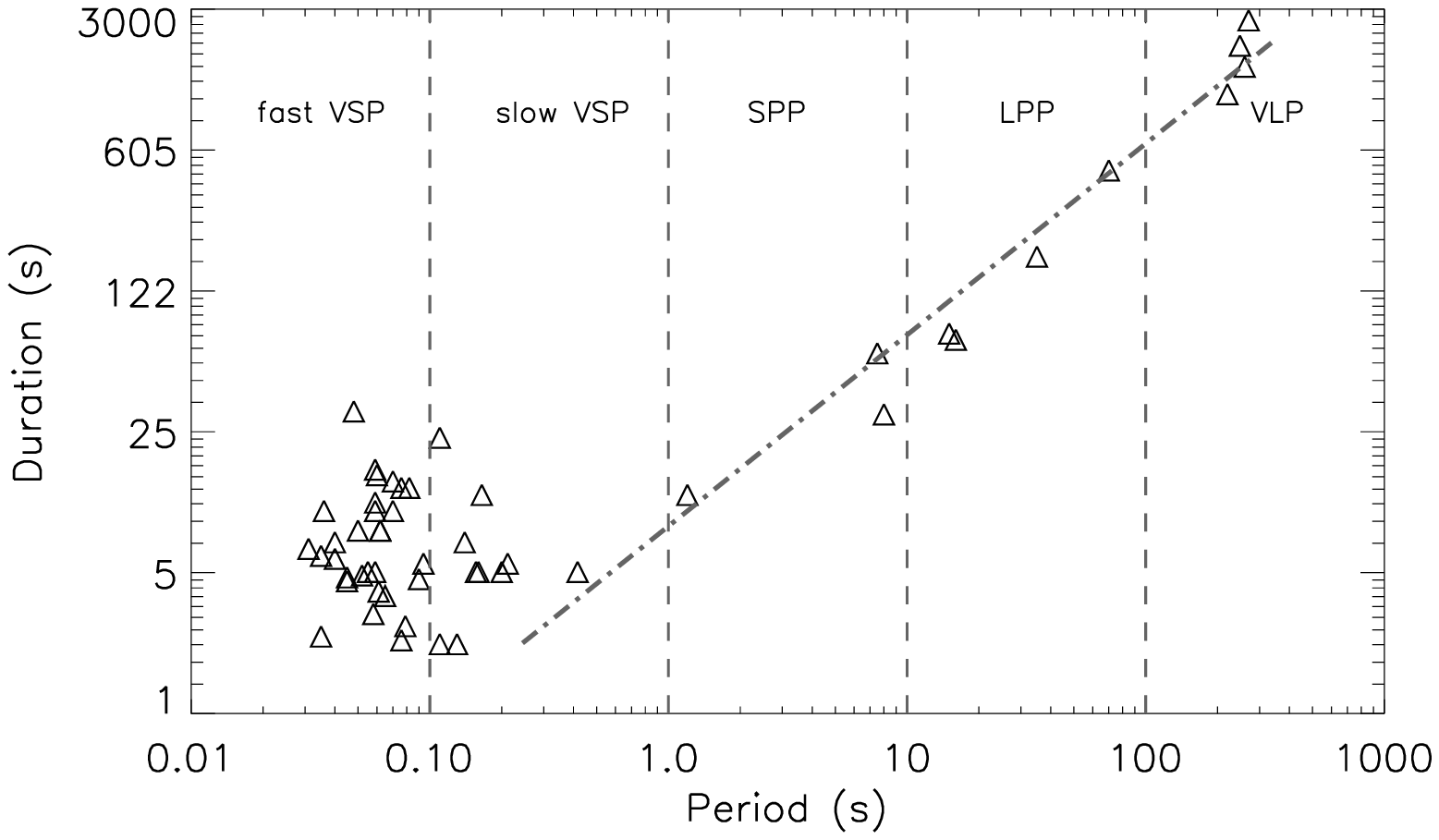}
   \caption{Relationship between the duration and the periods of QPPs.
   Each triangle denotes one QPP case, and the dot-dashed line is linear fitted.}
\end{center}
\end{figure}

\begin{table}\def~{\hphantom{0}}
  \begin{center}
  \caption{A brief summary of QPP with different timescales. Max-F indicates the magnitude of pulsating emission flux at each pulse in QPP}
  \label{tab:kd}
  \begin{tabular}{lccccccccccccc}\hline
   Hierarchy          &        VLP       &   LPP            &     SPP           &  slow-VSP        & fast-VSP        \\\hline
   D(s)               &  1140 $\sim$2640 &   90$\sim$450    &    20$\sim$80     & 2.2 $\sim$ 23    & 2.3 $\sim$ 16   \\
   Max-F(sfu)         &   90 $\sim$1200  &   70 $\sim$ 220  &    40$\sim$115    & 20 $\sim$ 95     & 15  $\sim$ 70   \\
   $b_{w}$(MHz)       &     $>1200$      &    $>1200$       &    $\sim$ 900     & 250 $\sim$ 700   & 220 $\sim$ 400  \\
   Mean period(s)     &  220$\sim$270    &   15 $\sim$70    &    1.2 $\sim$ 8   & 0.11 $\sim$ 0.416& 0.021$\sim$ 0.09\\
   r($\%$)            &  8 $\sim$ 45     &   10 $\sim$ 20   &    45 $\sim$ 65   &  45 $\sim$ 60    & 55  $\sim$ 75   \\
   Q-factor           &   6 $\sim$ 10    &    5 $\sim$ 7    &     4 $\sim$ 10   &  12  $\sim$ 209  & 30 $\sim$ 275   \\
   $R_{spdf}$(MHz/s)  &  -440 $\sim$ 389 & -610 $\sim$ 78.5 & -4000 $\sim$ 3500 & 6500$\sim$20000  & 8330$\sim>$30000\\\hline
   Cases              &        4         &         4        &          3        &        10        &      $>$ 30     \\\hline
  \end{tabular}
 \end{center}
\end{table}

Table 2 presents a brief summary of VLP, LPP, SPP, slow-VSP, and
fast-VSP.

\section{Discussion of the Physical Mechanisms}

From above investigations, we find that there are 5 classes of
QPPs with different timescales associated with the flare/CME
event: VLP, LPP, SPP, slow- and fast-VSP. They form a broad
hierarchy of timescales of hecto-second, deca-second, a few
seconds, deci-second, and centi-second. And this is similar to the
discovery of microwave burst timescales by Kruger et al (1994).
They classified the microwave burst timescales as tens of minutes,
a few minutes, a few seconds and sub-seconds which represent to
main burst phase, main burst pulse, subpulse and spiky burst
elements, respectively. But, these bursts have no obvious
periodicities. However, our investigations indicate that the most
remarkable feature of QPPs is the quasi-periodicity of the
repetitive pulses. Then, what is the generation mechanism of the
above various classes of QPPs?

It is well known that harmonic motion of a classic mechanic
oscillator is due to a restoring force proportional to the
displacement, and any elastic body can be excited to oscillate in
eigen-modes. When a magnetic field presents in plasma systems, the
characteristic eigen-frequencies will be determined by the
magnetic field, plasma density, temperature, and geometrical
configurations of the magnetized plasma system. Aschwanden (1987)
presented an extensive review of pulsation models and classified
them as three groups:

(1) MHD flux tube oscillations, which modulate the radio
emissivity with a standing or propagating MHD waves, e.g. slow
magnetoacoustic mode, fast kink mode, fast sausage mode (Roberts,
Edwin \& Benz, 1984; Nakariakov \& Melnikov, 2009).

(2) Periodic self-organizing systems of plasma instabilities of
wave-particle or wave-wave interactions interlocked by a
Lotka-Volterra type of coupled equations (Aschwanden \& Benz,
1988).

(3) Modulation of periodic acceleration (repetitive injection of
particles into the emission source region) which may possibly
generated from repetitive magnetic reconnections, for example, the
pulsed acceleration in solar flare (Aschwanden et al, 1994;
Aschwanden, 2004) or the multi-scale cascading reconnection
processes in the current sheet (Kliem et al, 2000; Karlicky et al,
2005).

\subsection{VLP Mechanism}

At first, we investigate the generation mechanism of VLP. From
Table 1 we know that periods of VLPs are 220 -- 270 s, they are
very similar to the oscillations obtained by Aschwanden et al
(1999) from the TRACE 171\AA observations. In that case the period
of oscillations is 276 s, and they interpret them as a standing
fast kink mode and might be a resonant coupling with the
photospheric 5-min p-mode oscillations. The period is expressed
as:

\begin{equation}
P_{kink}^{fast}=\frac{2L}{sc_{k}}=\frac{4\pi^{1/2}L}{s}(\frac{\rho_{o}+\rho_{e}}{B_{0}^{2}+B_{e}^{2}})^{1/2}\simeq6.48\times10^{-17}\frac{L\sqrt{n_{e}}}{B}.
\end{equation}

Here, the number of nodes is $s-1$,
$c_{k}=(\frac{\rho_{0}v_{A}^{2}+\rho_{e}v_{Ae}^{2}}{\rho_{0}+\rho_{e}})$
is the mean Alfv\'en speed for the inhomogeneous medium, $L$ is
the loop length (m), $\rho=n_{i}m_{i}$ is the plasma density and
subscript o and e refer to inside and outside of the loop,
$n_{i}\sim n_{e}$ (m$^{-3}$), the magnetic field $B$ is in Tesla.

\begin{figure}                        
\begin{center}
  \includegraphics[width=8.5cm]{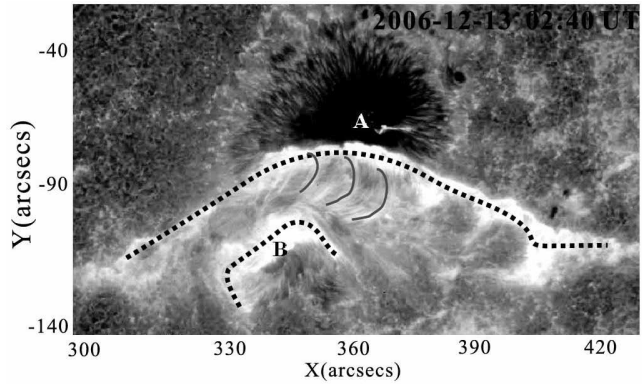}
   \caption{Skeleton map of the active region AR 10930 observed by Ca II H line on SOT/Hinode around the flare peak (2006-12-13, 02:40 UT).
   The thick black dotted lines indicate the flare ribbons, and the red solid lines sketched part of the small bright flux loops. A is the leading sunspot,
   and B is the following rotating sunspot.}
\end{center}
\end{figure}

In our case, the active region AR 10930 is an isolated one on the
solar disk during about 20 days before and after the flare event
(Fig.13). And it is structured mainly in two big sunspots, the
leading sunspot (marked as A in Fig.13) is about $50''$ (about
36,000 km), and the diameter of the following sunspot (marked as B
in Fig.13) is about 15,000 km. Over these two sunspots there are
many loops which connect from one sunspot to the other. We may
assume that the foot-points are located near ribbons (the thick
black dotted lines in Fig.13). Then we can estimate the lengths of
loops. The distance between two ribbons is in the range of $19.5''
- 39''$, that is about 14 - 28 Mm. Suppose loops are semicircles,
the loop lengths are $L= 44 - 88$ Mm. The plasma density and
magnetic field strength of the emission source region are
estimated as $B\sim 50 - 200$ G, $n_{e}\sim 10^{11}$ cm$^{-3}$
(Yan, et al, 2007). Based on these parameters and Equ.(1) we
obtain that the period is in the range of 9.0 - 36 s. Even if we
suppose the estimation with variation by a factor of 4, it is also
in the range of 2.5 - 144 s, too short to fit the observed periods
of VLP. Periods of standing fast sausage mode or propagating MHD
mode are shorter than that of standing fast kink mode, they are
not possible to be the candidate mechanism of VLPs.

Then what about the standing slow MHD modes? We know that both of
the standing slow sausage mode and standing slow kink mode have
periods:

\begin{equation}
P_{slow}=\frac{2L}{sc_{T}}=\frac{2L}{sc_{0}}[1+(\frac{c_{0}}{v_{A}})^{2}]^{1/2}\approx1.30\times10^{-2}\frac{L}{\sqrt{T_{e}}}.
\end{equation}

Here $c_{T}=c_{0}v_{A}/(c_{0}^{2}+v_{A}^{2})^{1/2}$ is the tube
speed in magnetic flux tube, $c_{0}$ is the sound speed, $v_{A}$
the Alfv\'en speed, $T_{e}$(K) is the temperature in the plasma
loop. We may use the brightness temperature to replace the plasma
temperature approximately:

\begin{equation}
T\sim
T_{B}=\frac{c^{2}D^{2}F}{k_{B}f^{2}r^{2}}\approx1.46\times10^{40}\frac{F}{f^{2}r^{2}}.
\end{equation}

Here $c$ is the speed of light, $D$ is the distance between the
Sun and the Earth, $F$ is the microwave emission flux intensity in
unit of sfu, $r$ is the radius of the emission source region in
unit of m. As the microwave flux intensity ($F$) is from the whole
sun, which may include a multiple loop sources. So, we may suppose
that the emission source of $F$ will be reasonably as the size of
whole active region, i.e. the size of AR 10930, which will be
about $r=2\times 10^{5}$ km. When $f=3.2$ GHz, the mean microwave
flux intensity $F=300$ sfu, then $T\sim 1.22\times10^{7}$ K. Then
we may gain the period is 186.4 - 372.6 s, which is very close to
the observed result of VLPs. So, we suggest that VLPs are possibly
generated from standing slow MHD modes.

Fig. 4 and Table 1 indicate that the magnitude of the pulsating
emission flux of VLP is in the range of 90 - 1200 sfu, which is
very strong modulation. In fact, each pulse of the VLPs is very
like a small eruptive process. We know that the kink mode can
change the magnetic field by altering the direction. But it can
change neither the magnitude of the magnetic field, nor the plasma
density of the loops directly. It is not clear how kink mode
produce so strong modulation of the microwave emissions. We would
rather believe that the VLP is generated by a standing slow
sausage mode.

On the other hand, Kosovichev and Sekii (2007) found that solar
umbra of the leading sunspot had a 3-min oscillation which started
immediately after the hard X-ray peak (50 - 100 keV in RHESSI) and
just before the soft X-ray maximum (GOES-12), with the amplitude
exceeding to that of pre-flare oscillations by a factor of 2 - 4.
However, at the same time Kosovichev and Sekii (2007) also pointed
out that the 3-minute umbra oscillation is uncertainty because of
the poorly observation cadence. As the periods of VLPs are 220 --
270 s, which are very close to the periods of the photospheric
5-min p-mode oscillations ($P=220 - 400$ s), we are apt to believe
that Kosovichev and Sekii (2007) underestimated the periods of the
umbra oscillations of the leading sunspot. The authors prefer to
deduce that VLP is a result of standing slow sausage mode coupling
and resonating with the underlying photospheric 5-min p-mode
oscillations. The modulation is amplified and constructs the main
framework of the whole flare/CME eruptive processes.

\subsection{LPP and SPP Mechanism}

Fig.12 indicates that almost all of VLPs, LPPs, SPPs, and part of
slow-VSPs are distributed around a line in the logarithmic
period-duration space, they are in a same QPP group. It is also
reasonable to suppose that LPP, SPP and part of slow-VSP have a
generating mechanism similar to that of VLP, at least they are
also generated from some MHD modulations. From Equ.(1) we get the
period in the range of 2.5 - 144 s, which is consistent with the
observed results of LPPs. Hence the standing fast kink mode may be
the most possible candidate mechanism of LPP.

As for SPP, the standing fast kink mode might be a weakly
candidate. The standing fast sausage mode should be the much
preferential mechanism. Its period is:

\begin{equation}
P_{sausage}^{fast}=\frac{2\pi
a}{c_{k}}\approx2.02\times10^{-16}\frac{a\sqrt{n_{e}}}{B}.
\end{equation}

$a$ is the loop width (m). From Fig.13 we may estimate the loop
width as about $1'' - 3''$, i.e. 700 - 2000 km. Then the period is
2.2 - 25.6 s. If the estimation has a variation by a factor of 4,
it is in the range of 0.6 - 100 s, which includes the whole range
of SPP and possibly LPP.

Actually, the propagating MHD mode have a period of a factor of
0.6 shorter than that of standing fast sausage mode. With the
parameters in our case, the period is in the range of 0.4 - 60 s.
Hence the propagating MHD mode is also a possible mechanism of LPP
and SPP.

Additionally, from Equ.(1) and Equ.(4) we may find that the
periods are most strongly dependent on the scales of the loop
(length or section radius). AR 10930 is a complex active region,
it is most possible that there is a variety of plasma loops of
which the scales are smaller than the above estimation. With such
smaller scale plasma loops, the periods of the standing fast kink
modes or the standing fast sausage modes are most likely to extend
to second and even sub-second, which may explain the mechanism of
LPP, SPP, and part of slow-VSP.

There is another alternative mechanism for the QPP with periods
$P>1$ s. Zaitsev et al (1998, 2000) proposed that a coronal loop
can be twisted and then carries an electric current. This
current-carrying plasma loop becomes a LRC-circuit resonator, and
the circuit oscillations can cause periodic modulation of loop
magnetic field, energy release rate and electron acceleration,
therefore, the emission of non-thermal electrons (Khodachenko et
al, 2005; Khodachenko et al, 2009). The eigen oscillation of the
LRC-circuit resonator may be a possible mechanism of microwave
QPP. When the electric resistance can be neglected, the period of
the circuit oscillations is:

\begin{equation}
P_{LRC}=\frac{2\pi}{c}\sqrt{\L\c{C}}\sim\frac{10^{12}}{I_{\varphi}}.
\end{equation}

Here, $\L=4l(log \frac{8l}{a\pi}-\frac{7}{4})$ is the loop
inductance, $\c{C}=\frac{c^{4}\rho S^{2}}{2\pi lI_{\varphi}^{2}}$
is the effective loop capacitance. $S=\pi a^{2}$, $a$ is the loop
section radius (m), $l$ is the loop length (m), $\rho$ is the
plasma mass density. $I_{\varphi}$ is the longitudinal electric
current (A) in the loop. The LRC-circuit model has been applied to
explain the microwave pulsation with periods of 0.7 -- 17 s
(Zaitsev et al, 1998). Observations show that the longitudinal
electric current is possibly in the range $2.87\times10^{10}$ --
$3.462\times 10^{12}$ A (Tan, 2007, etc), then we obtain the
oscillation period is about 0.3 -- 35 s, which is due to LPP or
SPP, and part of slow-VSP.

\subsection{VSP Mechanism}

Different from VLPs, LPPs, SPPs, and part of slow-VSPs (Group I),
the fast-VSP and the most part of slow-VSP are attributed to
another group (Group II), dispersively distributed away from the
line (Fig.12). This implies that the generation mechanisms of
fast-VSP and part of slow-VSP are intrinsically different from
Group I. In the work of Tan et al (2007) and Tan (2008), VSP is
explained as a result of modulations of the resistive tearing-mode
oscillations in some electric current-carrying flare loops, and
the pulsating emission is explained as plasma emission. The period
is:

\begin{equation}
P_{tear}=\frac{4\pi^{2}a^{2}\sqrt{\rho}}{I_{\varphi}\sqrt{M\mu_{0}}}\approx1.44\times10^{-9}\frac{a^{2}}{I_{\varphi}}\sqrt{\frac{n_{e}}{M}}.
\end{equation}

Here, $M$ is a parameter related to the distribution of the
electric current in plasma loop and the mode numbers of the
tearing-mode perturbations. $\mu_{0}$ is the permeability of free
space. From Fig.6 of Tan et al (2007), we know that $P_{tear}$ can
produce almost all kinds of VSPs with period of sub-seconds.

Additionally, the duration of quasi-periodic pulsating structure
is $D\simeq0.1513(\frac{\Delta
'}{jm\dot{B}_{\theta}})^{2/3}t_{A}^{2/3}t_{r}^{1/3}$ (Tan, 2008),
here $\Delta'$ is the tearing mode instability factor, $m$ is the
mode number, $t_{A}$ the Alfv\'en time scale (unit of second),
$t_{r}$ the resistive diffusive time scale (unit of second), $j$
the current density (unit of A $m^{-2}$), and $\dot{B_{\theta}}$
is the radial derivation of the poloidal magnetic field. We may
obtain the duration is about 200-300 s, which is much longer than
the observed durations (2.2-23 s). The most possible reason is
that there are many loops participating in the eruptive processes
in the flaring region, and the pulsating structure is evident only
when the contribution coming from a single flare loop. If the
contribution comes from several loops, the pulsating structure
will become diffuse or disappear because of superposition or
interference. So, our observations of VSPs are always fractional
sections.

The duration is dominated by the current density distribution,
magnetic configuration and plasma resistivity, while the period of
pulsation is dominated by the total electric current, the loop's
geometrical parameters, and the distribution of current density in
the cross-section. At the same time, the pulsating emission is
localized in some regions with small size, for example, localized
around magnetic islands in flaring plasma loops. From the
bandwidth of the emission we may estimate the perturbation of the
plasma density. From the frequency drifting rate we may estimate
the motion of the plasma loop and the motion of the energetic
particles. Combining all these observable parameters, we may probe
almost all the physical conditions and their evolutions.

In a word, the long period pulsation is possibly associated with a
large scale magnetized-plasma loop, while the short period
pulsation is possibly associated with some small scale
configuration in solar corona. The very short period pulsation may
reflect some local information about flaring plasma loops.

On the other hand, as all extraordinary VSPs are occurred
essentially after the main burst of the flare/CME event, we may
adopt the load/unload model (Nakariakov and Milnikov, 2009) to
interpret them. In load/unload model, QPP is a side effect of
transient energy releases, and the period is determined by a
buildup of free energy, the mechanism of the energy release, and
the energy outflow rate. We may suppose that energy release
mechanism is magnetic reconnection which is instantaneous and
short-lived, and energy outflow rate is determined by the magnetic
configurations which does not change observably after the main
burst of the flare/CME event. Then the period of QPP will be
determined mainly by the buildup of free energy. Naturally, very
after the main burst of the flare/CME event, the source region
becomes exhausted. So, the buildup of the free energy will take a
long time, the energy release lasts much less time, and cause
$W_{p} \gg W_{g}$. The strongly right circular polarization
indicates that the emission source arises in some small places
with simplex magnetic configurations.

\section{Summary}

From investigations in this work, we obtain following conclusions:
the timescales of microwave QPPs are distributed in a broad range
from hecto-second (VLP), deca-second (LPP), a few second (SPP),
deci-second (slow-VSP) to centi-second (fast-VSP), associated with
the X3.4 flare/CME event of 2006 December 13 in active region AR
10930. These QPPs are occurred successively around this event and
form a broad hierarchy. The higher hierarchy of QPP has longer
periods and durations, higher magnitude of the pulse emission
fluxes, wider frequency bandwidth, and weaker circular polarized,
lower frequency drift rates; while the lower hierarchy of QPP has
shorter periods and durations, lower magnitude of the pulse
emission fluxes, narrower frequency bandwidth, and stronger
circular polarized, higher frequency drift rates.

In logarithmic period-duration space, VLP, LPP, SPP, and part of
slow-VSP are approximately distributed around a line which implies
that all of them have the similar generation mechanism. Fast-VSP
and the most part of slow-VSP depart far from the above line which
implies that they possibly have different mechanisms.

Estimations show that VLP is possibly resulted from the standing
slow sausage modes coupling and resonating with the underlying
photospheric 5-min p-mode oscillations. It may be associated with
evolutive behaviors of the solar internal structures. As VLPs have
the largest magnitude of emission fluxes, we suggest that the
modulations are amplified and form the main framework of the whole
flare/CME eruptive processes. From Equ.(2) and (3), we can
estimate the radius of the emission source region:

\begin{equation}
r\simeq9.3\times10^{21}\frac{P}{fL}F^{\frac{1}{2}}.
\end{equation}

Here, the period ($P$), radio central frequency ($f$) and radio
emission intensity ($F$) can be obtained from radio observations,
the loop length ($L$) can be estimated from the optical or other
imaging observations approximately. Then we may obtain the radius
of the radio emission source region ($r$) by adopting Equ.(7) even
if we have no radio imaging observations. By substituting the
parameters obtained in the above sections, we may get the radius
of the emission source region from VLP paragraph A, B, C to D is
from $2.04\times10^{5}$ km, $2.06\times10^{5}$ km,
$1.62\times10^{5}$ km to $0.92\times10^{5}$ km, i.e., the source
region is undergoing an evolutive process of expanding at first,
and then shrinking. However, as we have no radio imaging
observations at the corresponding frequencies, we do not know the
exact cites of the emission source region, the only thing we can
do is to adopt the averaged loop length in our above estimations,
which may have much uncertainties.

Similar to VLP, LPP and SPP (may include part of slow-VSP) are
also caused by MHD oscillations. However, their MHD modes may have
some differences. They may be related with the standing fast
sausage or kink modes. The propagating MHD modes and the
LRC-circuit resonation of current-carrying plasma loops are also
the possible candidates of the generating mechanism.

Fast-VSP and most part of slow-VSP are generated by a completely
different mechanism: the modulation of the resistive tearing-mode
oscillations in electric current-carrying flare loops. In this
mechanism, both, the period and duration of QPP are coupled with
the magnetic field, plasma density, electric current, and the loop
parameters. By using their relation, we may deduce the physical
conditions of the emission source region.

The timescale of periods of QPPs implies a limit on the pulsating
emission source size. Regardless of generating mechanism the
pulsating source must be smaller than that given by the product of
speed of light and period ($P$). If not, the pulsating structure
would be smeared out (Elgar$\phi$y, 1986). So, it is reasonable to
suppose that the short periodic QPP may come from a smaller source
region. The broad hierarchy of timescales of QPPs occurred in a
flare event may imply that there is a multi-scale hierarchy of
sizes of the magnetic configurations in the flaring region, and
timescales of the dynamic processes. The frequency drift rate and
the bandwidth of the pulsating emission are dominated by the
emission mechanism which is always related to the magnetic field
strength, plasma density, and possibly to the plasma temperature.
It may be reasonable to suppose that the frequency drift features
of QPPs implies the motion of the pulsating source regions, and
the bandwidth of the pulsating emission are related to the
dimensional size of the pulsating source regions.

The period ratio between different classes of QPP have no obvious
trend. This fact may imply that there is no originated link
between different classes of QPP, even if they are occurred
simultaneous in the same frequency range. Actually, it is possible
that the short periodic QPP (e.g. fast-VSP, etc) is a small
quasi-periodic perturbation which superposed on the longer
periodic QPPs (e.g. VLP, etc), and the latter may dominate the
whole evolution of the flaring processes.

However, so far, because of the lack of imaging observations with
spatial resolutions in the corresponding frequency range, there
are many unresolved problems of the QPPs, for example, the spatial
behaviors, the spatial scales of the source region, etc. To
overcome such problems, we need some new instruments, for example,
the constructing Chinese Spectral Radioheliograph (CSRH, 0.4 - 15
GHz) in the decimetric to centimeter-wave range (Yan et al, 2009)
and the proposed American Frequency Agile Solar Radiotelescope
(FASR, 50 MHz - 20 GHz) (Bastian, 2003). Maybe, when these
instruments begin to work, we can get more and more cognitions of
the solar eruptive processes.

\acknowledgments

The authors would like to thank the referee's friendly and
valuably comments on the paper. Baolin Tan's work is supported by
NSFC Grant No. 10733020 and 10873021, Yin Zhang's work is
supported by NSFC Grant No. 10903013. This work is also partly
supported by MOST Grant No. 2006CB806301 and CAS-NSFC No.
10778605.

\end{document}